\documentclass[manuscript]{acmart}

\usepackage{graphicx}
\usepackage{subcaption}

\usepackage{todonotes}
\let\xtodo\todo
\renewcommand{\todo}[1]{\xtodo[inline,color=orange!75]{#1}}

\usepackage{enumitem}
\newlist{questions}{enumerate}{2}
\setlist[questions,1]{label=RQ\arabic*.,ref=RQ\arabic*}
\setlist[questions,2]{label=(\alph*),ref=\thequestionsi(\alph*)}
\AtBeginDocument{%
  \providecommand\BibTeX{{%
    \normalfont B\kern-0.5em{\scshape i\kern-0.25em b}\kern-0.8em\TeX}}}

\copyrightyear{2025}
\acmYear{2025}
\setcopyright{rightsretained}
\acmConference[MuC '25]{Mensch und Computer 2025}{August 31-September 03, 2025}{Chemnitz, Germany}
\acmBooktitle{Mensch und Computer 2025 (MuC '25), August 31-September 03, 2025, Chemnitz, Germany}
\acmPrice{}
\acmDOI{10.1145/3743049.3743057}
\acmISBN{979-8-4007-1582-2/25/08}

\newcommand{\lastaccess}{-- last accessed 2025-04-10}

\begin{document}


\title[The Impostor is Among Us]{The Impostor is Among Us: Can Large Language Models Capture the Complexity of Human Personas?}


\author{Christopher Lazik}
\affiliation{%
  \institution{HU Berlin}
  \city{Berlin}
  \country{Germany}}
\email{lazikchr@hu-berlin.de}

\author{Christopher Katins}
\affiliation{%
  \institution{HU Berlin}
  \city{Berlin}
  \country{Germany}}
\email{christopher.katins@hu-berlin.de}

\author{Charlotte Kauter}
\affiliation{%
  \institution{HU Berlin}
  \city{Berlin}
  \country{Germany}}
\email{charlotte.kauter@student.hu-berlin}

\author{Jonas Jakob}
\affiliation{%
  \institution{HU Berlin}
  \city{Berlin}
  \country{Germany}}
\email{jonas.jakob@student.hu-berlin.de}

\author{Caroline Jay}
\affiliation{%
  \institution{School of Computer Science, The University of Manchester}
  \city{Manchester}
  \country{United Kingdom}}
\email{caroline.jay@manchester.ac.uk}

\author{Lars Grunske}
\affiliation{%
  \institution{HU Berlin}
  \city{Berlin}
  \country{Germany}}
\email{grunske@informatik.hu-berlin.de}

\author{Thomas Kosch}
\affiliation{%
  \institution{HU Berlin}
  \city{Berlin}
  \country{Germany}}
\email{thomas.kosch@hu-berlin.de}

\renewcommand{\shortauthors}{Lazik et al.}

\begin{abstract}

Large Language Models (LLMs) created new opportunities for generating personas, expected to streamline and accelerate the human-centered design process. Yet, AI-generated personas may not accurately represent actual user experiences, as they can miss contextual and emotional insights critical to understanding real users' needs and behaviors. This introduces a potential threat to quality, especially for novices. This paper examines the differences in how users perceive personas created by LLMs compared to those crafted by humans regarding their credibility for design. We gathered ten human-crafted personas developed by HCI experts according to relevant attributes established in related work. Then, we systematically generated ten personas with an LLM and compared them with human-crafted ones in a survey. The results showed that participants differentiated between human-created and AI-generated personas, with the latter perceived as more informative and consistent. However, participants noted that the AI-generated personas tended to follow stereotypes, highlighting the need for a greater emphasis on diversity when utilizing LLMs for persona creation.

\end{abstract}

\ccsdesc[500]{Human-centered computing~Human computer interaction (HCI)}
\keywords{Personas, Large Language Models, User-Centered Design, Diversity}

\begin{teaserfigure}
  \includegraphics[width=\textwidth]{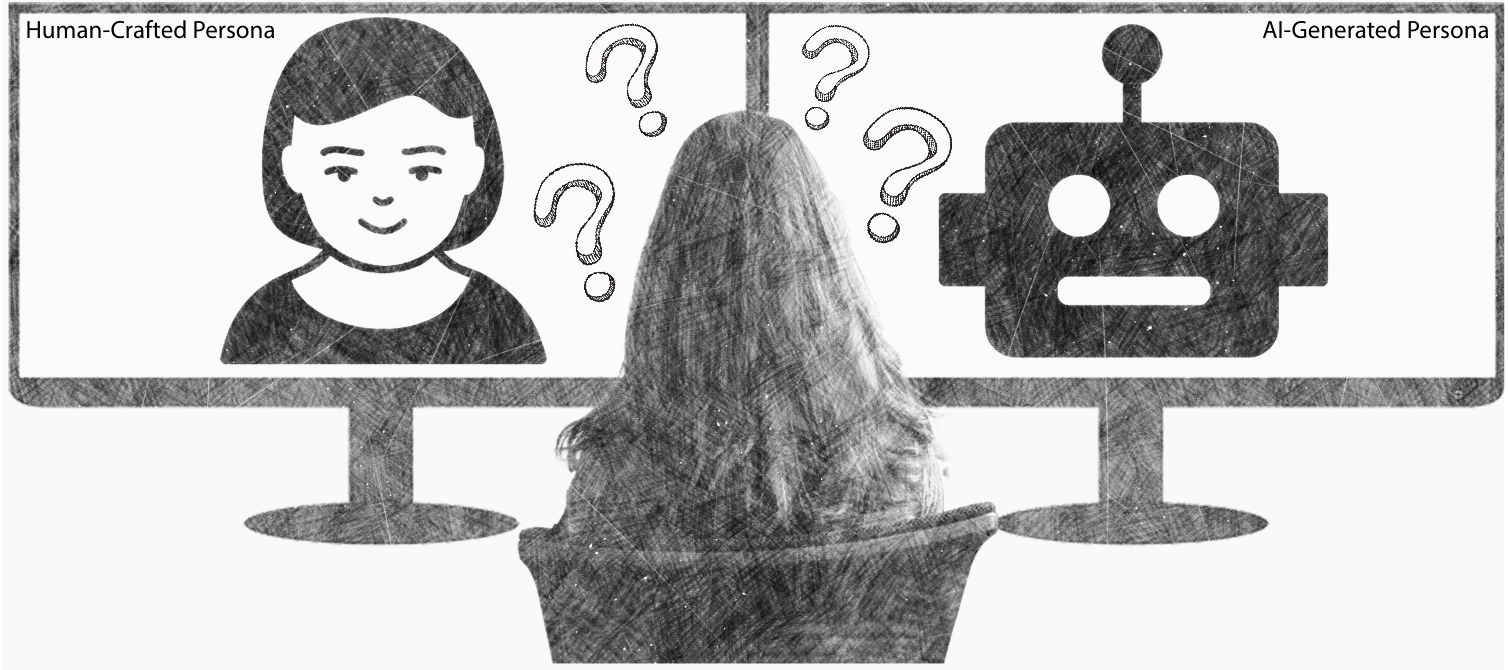}
  \caption{We investigated if and how users discern between textual descriptions of human-crafted and AI-generated personas. In a survey study, we investigated how common descriptions of personas, based on related work, affect the perceived realism and complexity of personas. We find that users can distinguish between human-crafted and AI-generated personas to a large extent, showing that stereotypicality, realism, and appeal are indicative features.}
 \Description{A grayscale sketch shows a person sitting at a desk, viewed from behind, looking at two computer screens. The screen on the left displays a simple illustration of a human face, labeled “human-crafted persona,” while the screen on the right shows a robot face, labeled “AI-generated persona.” Between the two screens, there are multiple question marks, symbolizing confusion or uncertainty about the two personas.}
  \label{fig:teaser}
\end{teaserfigure}


\maketitle

\section{Introduction}
Personas represent fictional characters created based on user research to represent different user types that might similarly use a service, product, site, or brand. Personas encapsulate key information in a narrative-written style~\cite{nielsen2013personas, Nielsen2019} including demographic details, behaviors, goals mostly using text~\cite{nielsen2015template}, and challenges of real users~\cite{Cooper1999}, helping Human-Computer Interaction (HCI) researchers keep the end user's requirements in mind during the design process~\cite{chang_personas_2008}. They support researchers and designers in better understanding and predicting the hypothetical users' needs, motivations, and frustrations, leading to more user-centered and effective design solutions. Subsequently, using personas ensures that the final research prototype or product is tailored to meet the needs of its intended audience, enhancing usability and user satisfaction~\cite{miaskiewicz_personas_2011}. However, creating appropriate personas can be time-consuming, thus slowing down the design phase and becoming a costly process~\cite{jansen_how_2022}.

Consequently, the automatic generation of personas based on data became interesting in the HCI community. For example, previous research mentioned that data-driven personas could be automatically generated using data from social media~\cite{9354402}, thus streamlining the persona generation process using crowd-sourced data. Recently, Large Language Models (LLMs) gained attention for simulating data of human participants in HCI research~\cite{aher_using_2023, argyle_out_2023, byun_dispensing_2023, chiang_can_2023, dillion_can_2023, gerosa_can_2024, gilardi_chatgpt_2023, hamalainen_evaluating_2023, heyman_impact_2023, horton_large_2023, park_social_2022, wang_want_2021} and as a data analysis tool in HCI research~\cite{tabone_using_2023}, with LLMs subsequently becoming interesting for persona generation~\cite{salminen_deus_2024,schuller_generating_2024}. LLMs can be used to create detailed personas, thanks to their ability to leverage their wealth of previously processed user data to generate realistic, contextually rich character profiles~\cite{schuller_generating_2024}. LLMs potentially allow for rapidly creating more comprehensive personas for various scenarios. Several AI-powered commercial tools and scientifc tools, such as PersonAI\footnote{\url{https://www.figma.com/community/plugin/1287786847239653675/personai-user-persona-generator} \lastaccess}, QoqoAI\footnote{\url{https://qoqo.ai/index.html} \lastaccess} or PersonaCraft~\cite{JUNG2025103445} exist to assist in generating personas. Thus, the HCI community became interested in examining if LLMs can create personas that represent the desired user base in terms of their efficacy and accuracy, as well as their potential biases and ethical implications~\cite{salminen_creating_2022, salminen_deus_2024, salminen_use_2022, schuller_generating_2024}, thus democratizing the complex persona creation process for novices and experts.

At the same time, researchers urge caution when using LLMs to generate user data~\cite{agnew_illusion_2024}. Hallucinations, value lock-ins, training bias, deceptive design patterns created by LLMs~\cite{krauß2024createfearmissingout, kosch_risk_2024, agnew_illusion_2024} may generate personas whose descriptions are not perceived as diverse or realistic. The existing research on using LLMs to generate personas faces two primary limitations. Previous studies have either not directly compared the perceptual differences between AI-generated personas and those crafted by humans using neutral persona descriptions~\cite{salminen_deus_2024}, or they have not thoroughly explored the key factors that affect how users discriminate between human-crafted and AI-generated personas~\cite{schuller_generating_2024}. In general, the quality of generated texts by LLMs can introduce a confirmation bias \cite{doi:10.1037/1089-2680.2.2.175}, especially to novices in a topic who have no expertise in persona creation. With the upcoming trend of using LLMs in persona generation, we see a potential threat to quality, specifically for those users who are not trained in human-centered design and seek fast solutions. Those novices might be misled by outputs that seem believable at first sight.

This paper addresses these limitations by conducting a user study to compare how users perceive human-crafted and AI-generated personas based on factors from previous research, including their informativeness for design, believability, stereotypicality, positivity, relatability, consistency, clarity, and likability~\cite{salminen_deus_2024, SALMINEN2020102437}. To this end, we surveyed participants to understand whether they can distinguish between human-crafted and AI-generated personas and which features in persona descriptions make them appear humane and credible. In the first step, we obtained a human-crafted persona dataset from HCI experts, who are aware of personas but do not frequently create them, to simulate a situation where non-experts develop personas. Then, we systematically created an AI-generated persona dataset. The consulted ten HCI experts crafted a persona based on factors relevant to persona design from previous work~\cite{burnett_gendermag_2016, salminen_deus_2024, SALMINEN2020102437}. We systematically prompted OpenAI's GPT-4o to generate personas~\cite{salminen_deus_2024}.  
In the second step, we presented the personas to 54 layperson participants who subjectively rated personas as to whether they were human-crafted or AI-generated and according to additional factors that characterize the credibility of personas. Our results show that our participants can discriminate between human-crafted and AI-generated personas. However, we also found that specific characteristics, such as stereotypicality and writing style, help people to distinguish between human-crafted and AI-generated personas. Our work shows that although LLMs convincingly generate personas, they do not entirely cover the complexity of users and are thus a potential threat to quality if chosen based on the wrong properties.

\section{Related Work}
We based our research on previous work concerning the relevance of personas and their data-driven creation in HCI. We present a review of the relevant literature in the following.

\subsection{Personas in Human-Computer Interaction}
Personas have different purposes in different scientific fields. For example, personas are developed in requirements engineering to express target users \cite{adlin_essential_2010}. In HCI, personas are fictional characters developed through user research to depict various types of people who may engage with a similar service, product, site, or brand. They convey essential details in a narrative format~\cite{nielsen2013personas, Nielsen2019}, covering demographics, behaviors, and goals—primarily through text~\cite{nielsen2015template}, as well as actual users' challenges~\cite{Cooper1999}, aiding Human-Computer Interaction (HCI) researchers in considering the needs of the end-users throughout the design process~\cite{chang_personas_2008}.

Designers can explicitly describe a fictional persona by encapsulating such information about potential user archetypes. Personas aim to offer an explicit common ground for developers to empathize with their users' needs without involving them directly in the design process \cite{salminen_use_2022}. Personas are also used in numerous other areas, such as defining potential customer groups in online marketing or in developing patient-oriented health and care technologies \cite{SALMINEN2020102437, 10.1145/3290605.3300880}. Personas are traditionally created manually by experts in the field of HCI and user experience design. Yet, this is a time-consuming process that requires not only a deep understanding of users but also specialized expertise in developing fictional and representative user profiles that are realistic \cite{schuller_generating_2024, schmidt_simulating_2024}. Creating realistic personas is an important balancing act since lowly detailed and less informative personas may not be picked up by practitioners~\cite{10.1145/2207676.2208573, 10.1145/1011870.1011884}.

Furthermore, the persona-creation process is influenced by the experience and perspectives of the experts involved, making it challenging to describe user groups in a truly representative manner, particularly for underrepresented users. Diversity and inclusion are central issues in persona creation. Diversity encompasses multidimensional differences in aspects of human beings (such as gender, age, ethnicity, neurodiversity, cultural background, and many more), while inclusion emphasizes actively involving various user groups to address their needs better. Failure to account for these aspects may result in products or systems that are not optimally usable for the broader user base. One approach specifically addressing gender diversity is the GenderMag method \cite{burnett_gendermag_2016}. These gender-specific differences affect how software is used and perceived. To mitigate these issues, the GenderMag approach uses fictional personas based on five cognitive dimensions: motivation, information processing style, computer self-efficacy, risk aversion, and tinkering. This helps identify potential usage barriers and fosters more inclusive design. In addition to addressing gender-related diversity, inclusion can be promoted through co-created personas. This approach directly engages users in the persona creation process, capturing their real needs and challenges to develop more accurate and practical personas. For example, in healthcare design, co-created personas have been used to reflect patients' experiences with Parkinson’s disease, dementia, or aphasia \cite{10.1145/3290605.3300880}. By incorporating their perspectives, designers can create solutions that better meet the unique needs of these user groups. Despite the promise of these methods, persona creation remains a time-intensive and complex process. Hard-to-reach populations, such as individuals with autism or special needs, are often underrepresented, making it difficult to develop personas that comprehensively reflect diverse user groups \cite{10.1145/3290605.3300880}.

\subsection{Automated Persona Generation}
Data-driven personas are profiles created using real data and analytics to represent key segments of users or customers, offering opportunities to enhance personalized marketing, product development, and user experience by making decisions based on concrete insights~\cite{Jansen2021, salminen2019future}. Considering the time and resources required to create personas~\cite{10.1145/3313831.3376502, 10.1145/997078.997089, 10.1007/978-3-642-03658-3_56}, automated approaches for persona creation have recently been proposed~\cite{Salminen2019}. In this context, LLMs offer promising potential. With their ability to process large amounts of text, LLMs could help close knowledge gaps and simulate the requirements of hard-to-reach user groups~\cite{schmidt_simulating_2024}.

A precise evaluation method is required to determine whether AI-generated personas realistically represent target groups. One potential tool is the persona perception scale~\cite{SALMINEN2020102437}, which provides a metric for measuring eight key dimensions, including credibility, consistency, and the willingness to use a persona. It enables the comparison of AI-generated and human-crafted personas to ensure that both types can be used effectively in the design process.
Salminen et al.~\cite{salminen_deus_2024} showed that users perceive AI-generated personas as realistic. The study found that these personas are primarily seen as consistent, credible, and informative \cite{salminen_deus_2024}. This highlights the potential of LLMs to generate personas perceived as believable, relatable, and informative while minimizing stereotyping, making them suitable for use in the design process \cite{salminen_deus_2024}.

In this context, Schmidt et al. \cite{schmidt_simulating_2024} emphasized the potential of LLMs in human-centered design. They argued that human resources should be used more efficiently and that tasks that can be handled just as well or even better by LLMs should be left to them. They also highlighted the importance of transparency: it has to be communicated whether and how LLMs are used in the design process to ensure trust and accountability \cite{schmidt_simulating_2024}. 

Agnew et al.~\cite{agnew_illusion_2024} warned that although LLMs have gained popularity for simulating human participants, they can rarely represent real users' profound experiences and perspectives. Important facets of human interaction, such as emotions, cultural contexts, or non-verbal communication, may not be captured by LLMs.  Kosch and Feger \cite{kosch_risk_2024} pointed out problems such as a lack of reproducibility, as LLMs do not consistently deliver the same results for a given input~\cite{kosch_risk_2024}. There is also a risk of bias, as LLMs are based on data that contains cultural and social biases, which can lead to skewed or stereotypical personas. Another risk is \textit{``value lock-in''}, where old norms and values are incorporated into the generated personas and prevent progress. It should be underlined that a transparent and responsible use of LLMs is necessary to recognize and minimize these risks early.

To improve the reproducibility and quality of AI-generated personas, \cite{salminen_deus_2024} presented a structured prompt strategy. This includes a clear definition of the desired persona characteristics such as age, gender, and occupation to guide the output of the LLM. Iterative customization of the prompts is used to refine the generated personas and ensure consistency. Finally, a review and expansion of the \textit{"skeletal"} personas is conducted to create complete narrative descriptions that are realistic and consistent. This strategy increases the reproducibility of the generated personas by ensuring that the inputs are aligned with the desired outcomes and that the personas are systematically improved. Additionally, the researchers suggest following three guidelines when implementing LLMs into the design process of personas: Verifying the personas regarding diversity and bias, verifying the personas using subject-matter experts, and adjusting the prompts if you observe challenges. To this end, Jung et al. \cite{JUNG2025103445} presented PersonaCraft, a tool to generate personas based on human survey data.

Schuller et al.~\cite{schuller_generating_2024} examined the perception of AI-generated and human-crafted personas and found no significant differences in perceived quality between the two types of personas. However, the study does not consider explicit factors that influence the perception of personas, such as those defined in the persona perception scale \cite{SALMINEN2020102437, salminen_deus_2024}, focusing instead on subjective impressions such as language and style \cite{schuller_generating_2024}. In fact, Shin et al.~\cite{10.1145/3643834.3660729} found that LLMs are not optimal at capturing key characteristics when generating personas. The authors suggested that humans should group data into persona-relevant categories beforehand and then use LLMs to summarize the data into personas.


In summary, personas are an essential asset for capturing the requirements in user interface design. Related work suggested accelerating the persona creation process through the support of LLMs~\cite{schmidt_simulating_2024}. Previous work found that LLMs capture the requirements of generated personas well~\cite{salminen_deus_2024} to the degree that users may not be able to distinguish between human-crafted and AI-generated personas~\cite{schulhoff_ignore_2023}. Yet, previous work did not investigate how common persona-driven features, such as informativeness, reliability, consistency, or clarity~\cite{salminen_deus_2024, SALMINEN2020102437}, are perceived differently between human-crafted and AI-generated personas. Our paper closes this gap by conducting a survey study investigating which features make personas appear humane or generated.
\section{Methodology}
Previous studies have demonstrated the significance of personas in user experience design. As a result, research has explored the potential of using LLMs to expedite the persona creation process, suggesting that including simulated user data could enhance the diversity of the generated personas. Nevertheless, the rise of LLMs might lead to the assumption of accelerating development processes by directly generating personas instead of crafting them.

Prior work has not examined whether users can distinguish between human-crafted and AI-generated personas based on common perceptual factors contributing to a persona's credibility. Consequently, this paper is influenced by earlier research that focused exclusively on AI-generated personas~\cite{salminen_deus_2024} or studies that compared human-crafted and AI-generated personas but provided limited evaluation of key perception variables~\cite{schuller_generating_2024} such as consistency, completeness, or clarity~\cite{SALMINEN2020102437}. Therefore, we define the following research questions:
\begin{itemize}
    \item[\textbf{RQ1:}] To what extent can users distinguish between human and AI-generated personas?
    \item[\textbf{RQ2:}] How do the perceived differences between human-created and AI-generated personas relate to specific features of the persona description?
    \begin{itemize}
        \item[\textbf{RQ2.1:}] What are the differences between human-crafted and AI-generated personas regarding quality aspects from the literature?
        \item[\textbf{RQ2.2:}] How do participants characterize the difference between human-crafted and AI-generated personas?
    \end{itemize}
\end{itemize}

We conducted one data preparation step and a persona-comparison study to answer the research questions. In the first step, we asked ten HCI experts to craft one persona each, resulting in ten human-crafted personas that were not part of any training set from an LLM. Then, we utilized a prompting strategy by Salminen et al.~\cite{salminen_deus_2024} to generate ten personas using an LLM. The user study directly compared the human-crafted personas to generated personas by exposing participants to personas and asking them to assess if a human or an AI created the persona. Furthermore, we employed several constructs used by Salminen et al.~\cite{SALMINEN2020102437, salminen_deus_2024}. We specifically focused on a subset of these dimensions most relevant to understanding the perceived realism, coherence, and usefulness of personas: informativeness, believability, stereotypicality, positivity, relatability, consistency, clarity, and likability. We omitted dimensions tied directly to designer-specific tasks, such as completeness and willingness to use~\cite{SALMINEN2020102437}, since our participant pool was not composed of active design professionals applying personas in real-world projects. Additionally, similarity and empathy were excluded because they involve personal relatability to the persona, which could introduce confounding variables unrelated to the primary research questions. We obtained ethical approval for the studies from the institutional review board of our university.



\section{Obtaining Human-Crafted Personas}
Our first data acquisition step was designed to provide a baseline of human-crafted personas. We created new personas by collaborating with researchers and practitioners with several years of HCI experience in the form of a survey. This approach ensures we have a controlled set of personas free from external context-specific biases, unlike those that might arise from using personas from an online resource that targets a specific use or has an intent. We want to investigate the appeal of personas instead of introducing a bias through particular use cases that might influence the appeal and freedom of the created personas. Furthermore, personas that were available online can be in the training corpus of LLMs that we used in our study to generate personas. By obtaining a new set of human-crafted personas, we ensured that we would compare them against AI-generated personas that were not part of the training set of any LLM before. We created an online survey to collect personas from HCI experts using persona properties recommended by the GenderMag project~\cite{burnett_gendermag_2016}\footnote{\url{https://gendermag.org/custom_persona.php} \lastaccess} and Salminen et al.~\cite{salminen_deus_2024}. We did not specify a persona complexity level for the participants, allowing creative freedom.

The choice of asking general HCI experts instead of more specific persona design experts ensures that our participants are familiar with the concept of personas while being less far from novices.

\subsection{Survey Structure}
The survey aims to obtain human-crafted personas that were previously not used to train LLMs or are the results of an LLM. Furthermore, the set of crafted personas should be comparable to personas made by knowledgeable novice users.  Our survey for collecting human-crafted personas consisted of two parts.
First, we explained the course of the survey to the participants and asked for informed consent. Second, we gave a template to our participants as shown in \autoref{table:template}, including the following features for a fictional persona. The properties considered included name, age, occupation, background and skills, motivation and strategies, technological attitude, and other details characterizing the persona. We chose these properties from the previous work by Salminen et al.~\cite{salminen_deus_2024} on generating personas using LLMs and completed the properties using the recommendations by the GenderMag project~\cite{burnett_gendermag_2016}. These attributes provide the basic information about a general persona. While their name, age, and occupation give a general context of the persona's circumstances, the attribute called ``details'' includes a description of the persona with more details to provide context. These general details allow the person using the persona to understand the general context of the described fictional person. To enable the usage of the personas in our study for software development processes, we additionally included background and skills, motivation and strategies, and attitude to technology. These attributes are based on the GenderMag concept. Aiming to address the general approach to challenges and technology, they are a fitting extension to the more general personas. 

We told all participants not to use generative AI, specifically LLMs, in their persona creation process. We also asked the participants to provide the information in sentences for the last four properties the participants should describe (i.e., background and skills, motivation and strategies, technological attitude, and other details). Even though personas might include a picture \cite{adlin_essential_2010}, we did not ask the participants to provide one to keep comparability to AI-generated personas. Furthermore, previous research showed that including pictures introduces biases within the individual assessment and perception of personas~\cite{10.1145/3173574.3173891}. The findings show that contextual photos enhance informativeness. At the same time, images of different people confuse, and user biases influence how personas are interpreted, suggesting a careful selection of photos in persona design. As a consequence, we refrained from using photographs in the persona descriptions. We conducted the survey using LimeSurvey\footnote{\url{https://www.limesurvey.org}\lastaccess}, hosted by our research institute.

\begin{table*}
\centering
\small 
\renewcommand{\arraystretch}{1.2} 
\caption{Survey template for participants who created human-crafted personas. Our participants filled in the fields that ultimately described the persona.}
\Description{Survey template for participants who created human-crafted personas.}
\label{table:template}
\begin{tabular}{p{0.3\textwidth} p{0.7\textwidth}} 
\toprule
\textbf{Attribute}                & \textbf{Explanation}                                                           
\\ \midrule
\textbf{Name}                      & The name of the persona you want to craft.                                                               
\\ 
\textbf{Age}                      & The age of the persona you want to craft.                                                               
\\ 
\textbf{Occupation}               & What is the persona you are creating doing for a living?                                                        
\\ 
\textbf{Background and Skills}              & What knowledge and skills does the persona that you are creating have? This may include educational qualifications or knowledge in tools. Feel free to include every skill or other background you think is part of the identity of your hand-crafted persona.              
\\ 
\textbf{Motivation and Strategies}     & What motivates your created persona during their everyday life? How do they approach challenges and tasks?    
\\ 
\textbf{Attitude to Technology}   & What does your created persona think about technology?                                       
\\ 
\textbf{Details} & Describe your created persona a little bit more to give more context of their personality. 
\\ \bottomrule
\end{tabular}
\end{table*}

\subsection{Procedure}
We greeted the participants and explained the survey's objectives to them. Following this, the participants completed an informed consent form. We informed the participants that their participation is entirely voluntary and their right to withdraw their participation and data anytime without any disadvantages from the survey. Participants were informed that they were not receiving compensation. We collected demographic information from the participants, including their age, self-identified gender, profession, research focus, experience with personas, and years of experience in HCI research or user-centered design. Participants were also asked to select a pseudonym, which could be used to request data deletion through a pseudonymization list. They were then provided with a comprehensive task description detailing the persona creation process, which included the specified characteristics of the persona. The properties were concerned with background and skills, motivation and strategies, technological attitude, and other details that describe the personality of the persona. The participants filled in the persona details without any constraints, except for the requirements to answer in complete sentences and avoid using generative AI. Lastly, they were allowed to add any additional comments to the survey.




\subsection{Participants}
We recruited ten participants (five self-identified as female, four self-identified as male, and one self-identified as non-binary) with ages ranging from 23 to 36 ($\bar x = 30.20, s = 3.39$). All of our participants had a background in HCI research with an average experience of 4.65 years ($s = 2.69$). While the concept of personas was clear to all ten participants, four reported creating personas multiple times during a year, one reported creating personas various times during a month, and five reported that they had never made a persona before. We explained the survey's purpose to the participants and informed them they could cancel it anytime.

\subsection{Results}
We collected a total of ten human-crafted personas. Two researchers rated and screened the set of personas regarding their plausibility. Then, both researchers independently coded the personas using the same coding sheet proposed by Salminen et al.~\cite{salminen_deus_2024}. The rating criteria are displayed in~\autoref{table:crafted_criteria}. We calculated Krippendorff's $\alpha$ as a measure for the agreement of the codes towards the criteria between both researchers. We found that Krippendorff’s $\alpha$ resulted in  $\alpha = 0.879$, meaning that sufficient agreement was reached\footnote{A Krippendorff's $\alpha$ of $\geq .800$ is considered high agreement between the coders.}. The age of the human-crafted personas ranged from 10 to 72 ($\bar x = 51.40, s = 19.26$). The HCI experts created seven male, two female, and one non-binary persona. The text length, counted in words, ranged from 71 to 433 words ($\bar x = 260.7, s = 109.75$). While the physical appearance was only described in two personas, the personality was expressed in every persona. The occupations of the human-crafted personas were specified as retiree, artist, unemployed, firefighter, pupil, faculty member, researcher and director, and gardener. Retirees were chosen multiple times since the age of some personas was relatively high. To ensure no impact of typographical errors in the human-crafted personas, we corrected spelling mistakes across all human-crafted personas. This is because typographical errors would reveal that a human created a persona for later comparison with AI-generated personas.

\begin{table*}
\centering
\small 
\renewcommand{\arraystretch}{1.2} 
\caption{Evaluation sheet for assessing the qualitative agreement of the human-crafted personas. The attributes of the evaluation sheet were adapted from previous work~\cite{salminen_deus_2024}.}
\Description{Evaluation sheet for assessing the qualitative agreement of the human-crafted personas.}
\label{table:crafted_criteria}
\begin{tabular}{p{0.2\textwidth} p{0.5\textwidth} p{0.25\textwidth}} 
\toprule
\textbf{Attribute}                & \textbf{Explanation}                                                            & \textbf{Code} 
\\ \midrule
\textbf{Age}                      & The persona's age                                                               & number 
\\ 
\textbf{Gender}                   & The persona's gender                                                            & 'f', 'm', 'nb', 'o'
\\ 
\textbf{Occupation}               & The persona's occupation                                                        & job title 
\\ 
\textbf{Text Length}              & The length of the persona's properties that were given in sentences              & word count 
\\ 
\textbf{Physical Appearance?}     & Is the persona description mentioning the persona's physical appearance?        & yes or no 
\\ 
\textbf{Personality Mentioned?}   & Is the persona's personality described?                                         & yes or no 
\\ 
\textbf{Informativeness for Design} & Does the persona description contain adequate information to design an app or system to address the persona’s needs? & yes or no 
\\ 
\textbf{Believability}            & Does the persona appear realistic, i.e., lifelike, like an actual person that could exist? & yes or no 
\\ 
\textbf{Stereotypicality}         & Does the persona appear stereotypical? (Stereotypes are related to a widely held but fixed and oversimplified image or idea of a particular type of person or thing.) & yes or no 
\\ 
\textbf{Positivity}               & Is the person depicted in a positive light?                                      & yes or no 
\\ 
\textbf{Relatability}             & Is the persona relatable? (Relatability is the quality of being easy to understand or feel empathy for.) & yes or no 
\\ 
\textbf{Consistency}              & Is the persona consistent? (A consistent persona does not have conflicting information e.g., the person is described as being overall happy but later as being sad in general.) & yes or no 
\\ \bottomrule
\end{tabular}
\end{table*}

\section{Evaluating the Perception Between Human-Crafted and AI-Generated Personas}
Using the personas from the first data acquisition step as a set of human-crafted personas, we conducted a second survey to determine the properties that distinguish between human and AI-generated personas. To directly compare human-crafted personas and personas created using an LLM, we asked participants to label a mixed set of generated and human-crafted personas. This section explains the persona-generation process, the survey methodology, and the results.

The evaluation of personas in our study was conducted with survey participants who were not explicitly designers to explore broader perceptions of realism, informativeness, and stereotypicality, important attributes linked to persona effectiveness. While personas are ultimately used by designers, understanding how non-designers perceive the authenticity and detail of persona descriptions provides foundational insights into their general credibility and appeal. This approach aligns with prior research that examines personas from a user-perception perspective to assess their believability, perceived utility, and credibility of traits~\cite{salminen2018personas}. By gathering input from individuals who could represent diverse perspectives, we want to identify generalizable traits that influence persona acceptance.

\subsection{Generating Personas using AI}
We used GPT-4o\footnote{\url{https://openai.com/index/hello-gpt-4o} \lastaccess} to generate personas, using a zero-shot structured prompting strategy from related work~\cite{salminen_deus_2024}. We used a zero-shot prompting strategy to emulate a typical real-world scenario where users generate personas with minimal customization or prior examples, reflecting the baseline capabilities of the LLM. To avoid the token limit given by GPT-4o\footnote{GPT-4o encounters a token limit of 4096 tokens.}, the approach begins with the generation of a skeleton. These skeletal personas include the personas' properties (e.g., name, age, occupation, and details). After creating a set of ten skeletal personas, we iterate over the persona skeletons and prompt GPT-4o to expand each skeletal persona into a complete one. This approach generates the main aspects of all the personas at once that are \textbf{structured} in a way that is \textbf{comparable} to both our human-crafted personas and other generated personas from the literature.
Furthermore, to achieve personas that provide information that can be \textbf{used} to develop software, we expanded the main properties with three additional elements (background and skills, motivation and strategies, and attitude to technology) from GenderMag \cite{burnett_gendermag_2016}. Thus, the structure of our personas asks for general information about a fictional person and their knowledge base, how they approach challenges, and how they deal with technology. The resulting generation script can be found in our supplementary material.

We used two prompts, one for each generation step:

\begin{itemize}
    \item[\textbf{I}] ``Generate 10 user personas. Provide the output in a json array, with each dict containing only the following keys: "index", "name", "age",  "occupation",  "background and skills", "motivations and strategies", "attitude to technology", "details";  Index starts at 1. Start your response with the open square bracket [''
    \item[\textbf{II}] ``Expand on the following summary persona. Ensure that all the information provided is used in your expanded persona. Don't include more or less properties. Stick to the structure that is given.'' + the JSON-file
\end{itemize}

We generated ten final personas for this study, which were generated in one round. We did not generate and evaluate multiple personas for a good fit for the study. First, previous work~\cite{salminen_deus_2024} showed that generated personas appear credible when only prompted once. Second, we did not intend to bias the provided AI-generated persons through a pre-selection since a realistic use case is the spontaneous generation of personas using an LLM without further screening.

\subsection{Persona Generation Results}
Similarly to the first data acquisition survey, two experts from our research team reviewed the generated personas. Again, we use Krippendorff’s $\alpha$ and find a high agreement with an $\alpha = 0.949$. 

The age of the generated personas ranged from 27 to 50 ($\bar x = 35.90, s = 7.37$). GPT-4o created five male, five female, and zero non-binary personas. During the skeletal persona generation, the gender of the personas was consistently alternating. The text length of the given sentences, in words, ranged from 145 to 254 words ($\bar x = 191.10, s = 34.07$). 
No physical appearance was described for any of the personas. Every persona was defined by personality. The occupations of the generated personas were specified as software engineer, digital marketer, product manager, UX designer, data analyst, HR manager, CEO, sales manager, technical support, and business analyst. Every generated persona had a unique occupation. However, all listed occupations fall within the white-collar category and are settled in the technology, business, and management sectors.

Regarding the criteria in the lower part of \autoref{table:crafted_criteria} (i.e., the last eight items), our coders mainly answered with ``yes''. The personas were informative, believable, positive, relatable, and consistent. However, all of them appeared to be stereotypical to our experts. Furthermore, our experts noticed that many personas are described as performing similar thoughts in their free time as they do in their jobs.

\subsection{Survey Structure}
The survey study consisted of two parts. First, the participants were informed of the study goals and provided informed consent. Second, we collected demographic information of our participants and assessed their AI literacy employing the ``Scale for the assessment of non-experts’ AI literacy'' \cite{laupichler_development_2023}. Then, we informed the participants about their task of rating a mixed set of AI-generated and human-crafted personas and asked them to rate our 20 randomly ordered personas. Then, participants performed the task. They rated several statements on a 7-point Likert scale\footnote{1: Strongly Disagree; 2: Disagree; 3: Somewhat Disagree; 4: Neutral; 5: Somewhat Agree; 6: Agree; 7: Strongly Agree.} and provided reasons for their ratings in a free text field. We compensated our participants with nine Pounds per hour. The participants knew that personas were either human-crafted or AI-generated but were unaware of which personas were human-crafted or AI-generated throughout the survey. The survey was created using LimeSurvey\footnote{\url{https://www.limesurvey.org/}\lastaccess}, hosted by our research institute.

\subsection{Task}

\begin{figure}
    \centering
    \begin{subfigure}[b]{0.49\textwidth}
        \centering
        \includegraphics[width=\textwidth]{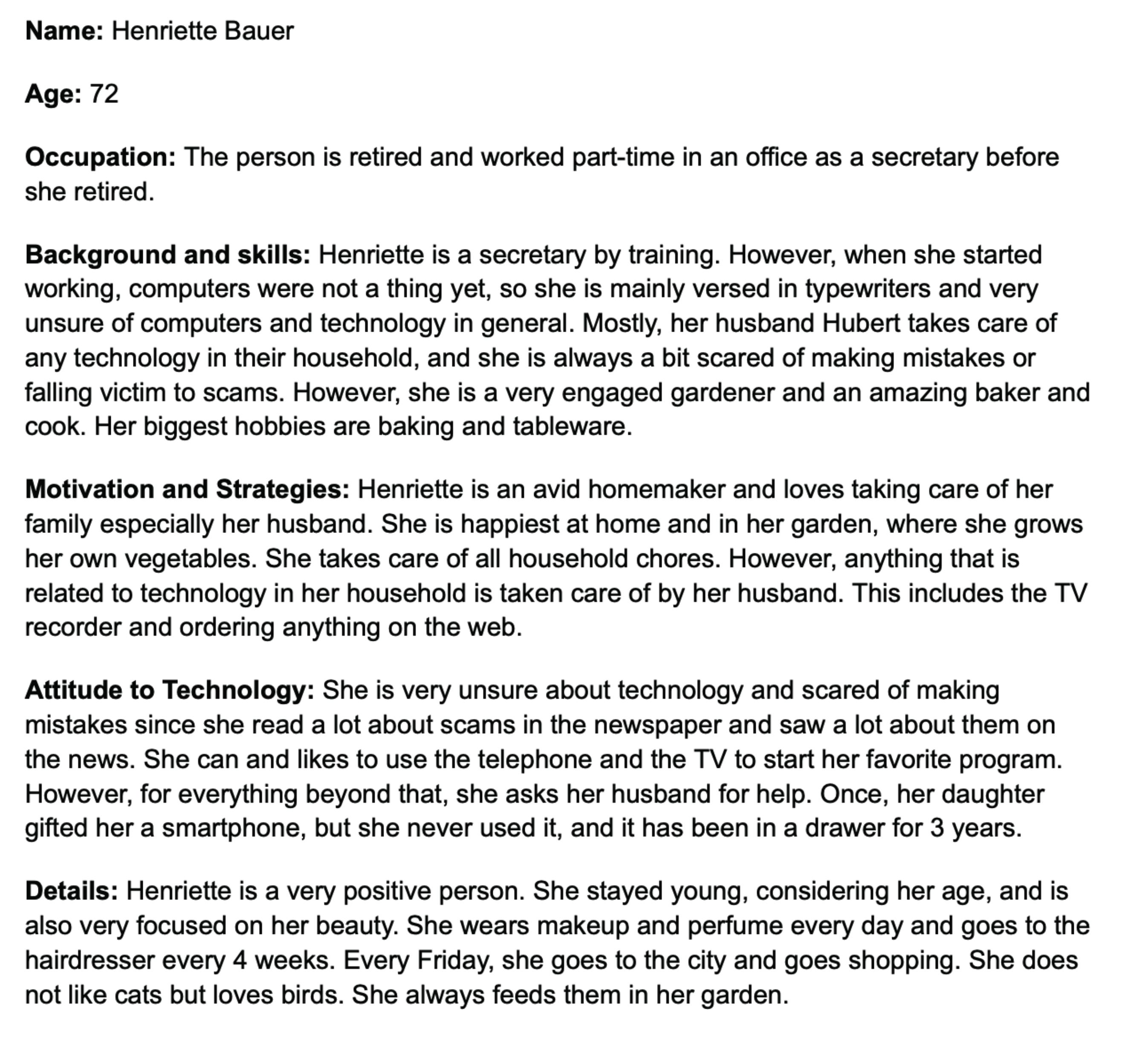}
        \caption{}
        \label{fig:humancraftedpersona}
    \end{subfigure}
    \hfill
    \begin{subfigure}[b]{0.49\textwidth}
        \centering
        \includegraphics[width=\textwidth]{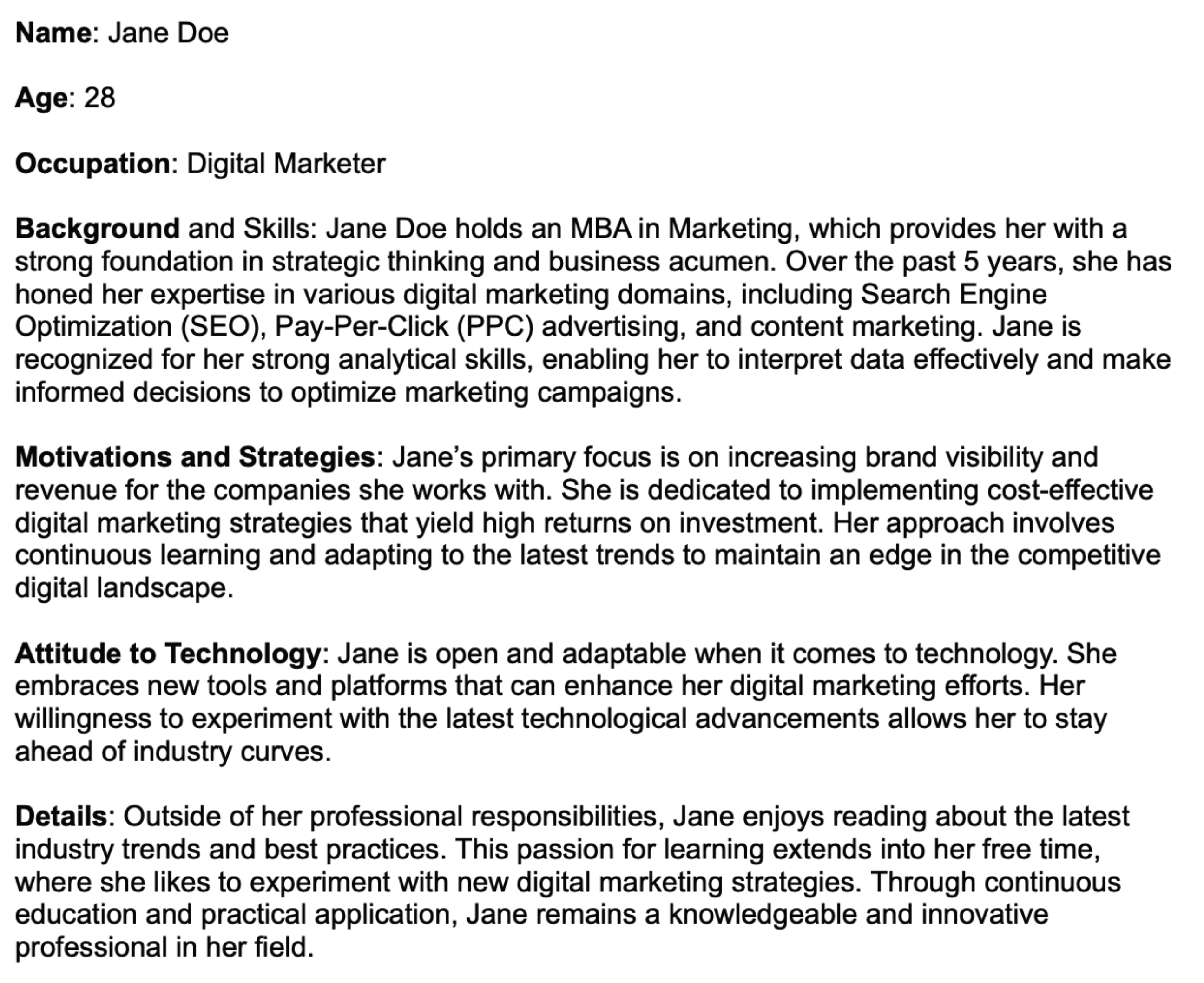}
        \vspace{4mm}
        \caption{}
        \label{fig:aigeneratedpersona}
    \end{subfigure}
    \caption{Two example personas from our survey study. \textbf{(a):} human-crafted. \textbf{(b):} AI-generated}
    \Description{A comparison of two personas presented side by side in paragraph format.}
    \label{fig:personas}
\end{figure}

In our survey study, the participants had to solve one main task. Each participant was exposed to a set of 20 personas. The order of these personas was randomized to prevent possible effects. Ten out of the 20 personas were generated using Open AI's GPT-4o. The other ten personas were handcrafted by experts from the field of HCI in our first data acquisition survey.  Participants were given the personas as a structured text description as shown in \autoref{fig:personas}. By displaying the personas in a neutral way, we aimed to avoid revealing whether a persona was human-crafted like shown in \autoref{fig:humancraftedpersona} or AI-generated as seen in \autoref{fig:aigeneratedpersona}. Participants were exposed to 15 statements that they had to rate on a 7-point Likert scale as shown in \autoref{table:aspects}.

\begin{table*}
\centering
\small 
\renewcommand{\arraystretch}{1.2} 
\caption{Statements ranked by participants to measure the persona aspects. Participants rated the persona aspects on a 7-point Likert scale.}
\Description{Statements ranked by participants to measure the persona aspects. Participants rated the persona aspects on a 7-point
Likert scale.}
\label{table:aspects}
\begin{tabular}{p{0.3\textwidth} p{0.7\textwidth}} 
\toprule
\textbf{Persona Aspect}                & \textbf{Statement}                                                           
\\ \midrule
\textbf{Human-Crafted}                      & The persona is human-crafted.                                                               
\\ 
\textbf{AI-Generated}                      & The persona is AI-generated.                                                               
\\ 
\textbf{Informativeness for Design}               & The persona description contains adequate information to design an app or system to address the persona’s needs.                                                      
\\ 
\textbf{Believability}              & The persona appears realistic, i.e., lifelike, like an actual person that could exist.              
\\ 
\textbf{Stereotypicality}     & The persona appears stereotypical. (Stereotypes are related to a widely held but fixed and oversimplified image or idea of a particular type of person or thing.)    
\\ 
\textbf{Positivity}   & The person is depicted in a positive light.                                       
\\ 
\textbf{Relatability} & The persona is relatable. (Relatability is the quality of being easy to understand or feel empathy for.) 
\\ 
\textbf{Consistency} & The persona is consistent. (A consistent persona does not have conflicting information e.g. the person is described as being overall happy but later as being sad in general.) 
\\
\\
\textbf{Clarity\textsubscript{1}} & The information about the persona is well presented. 
\\ 
\textbf{Clarity\textsubscript{2}} & The text in the persona profile is clear enough to read. 
\\ 
\textbf{Clarity\textsubscript{3}} & The information in the persona profile is easy to understand. 
\\
\\
\textbf{Likability\textsubscript{1}} & I find this persona likable. 
\\ 
\textbf{Likability\textsubscript{2}} & I could be friends with this persona. 
\\ 
\textbf{Likability\textsubscript{3}} & This persona feels like someone I could spend time with. 
\\ 
\textbf{Likability\textsubscript{4}} & This persona is interesting. 
\\ \bottomrule
\end{tabular}
\end{table*}

The first two statements were about the origin of the persona. By rating them, participants decided whether they think a persona is human-crafted or AI-generated. After that decision, participants rated multiple statements to assess the design, believability, stereotypicality, positivity, relatability, and consistency of the personas. These aspects of personas were based on the literature by Salminen et al.~\cite{salminen_deus_2024}. Additionally, we utilize further questions from the persona perception scale~\cite{SALMINEN2020102437}. We used the constructs ``Clarity'' to determine if the persona description was clear to the participants. Furthermore, we measure the construct ``Likability'' to further assess how personas appeal to participants in our survey. We did not include the constructs ``Consistency'' and ``Credibility'' since the constructs contained items asking for a profile picture. Yet, none of our personas contained a picture to avoid biases regarding the relatability~\cite{10.1145/3173574.3173891}. We excluded ``Completeness'' and ``Willingness to Use'' since they ask for items regarding a target group (e.g., managers or designers). Furthermore, we excluded ``Similarity'' and ``Empathy'' since both ask for items personally tied to the participant and their relation to the persona. We asked each participant why they assessed a persona as human-crafted or AI-generated and if they were undecided or conflicted about their choice. After rating the above-mentioned items of a persona, the participants were asked in an open text field what aspects of the persona influenced their decision. 

\subsection{Procedure}
The participants were introduced to the task after agreeing to participate in our study. Participants were informed about their task of rating generated and handcrafted personas. The participants had to iterate over twenty personas in a randomized order. After rating all the personas in our set, the participants were asked to provide qualitative feedback on the study experience to collect information on how participants perceived the task over the experiment. In the last step, we debriefed the participants, and they got confirmation that they finished the questionnaire.

\subsection{Participants}
We sampled participants through the online platform Prolific\footnote{\url{https://www.prolific.com} \lastaccess}. We recruited participants until we obtained 54 participants who passed two validation questions.

We collected the answers of 54 participants (25 self-identified as female, 27 self-identified as male, and two self-identified as non-binary) aged 18 to 73 ($\bar x = 32.04, s = 10.51$). The sample was heterogenous regarding the occupation of the participants. The participants reported using LLMs in different frequencies (11 ``Never'', 10 ``Multiple times during a year'', 15 ``Multiple times during a month'', 11 ``Multiple times during a week'', and 7 ``Every day'').
Participants responded to the statement ``I am familiar with the concept of personas.'' on a 7-point Likert scale. The mean response was 5.5 ($s = 1.28$), indicating that most respondents agreed with the statement. Additionally, we asked the participants to rate the statement ``I am familiar with crafting personas.'' on a 7-point Likert scale. Here, the average response was 4.39 ($s = 1.66$), which shows that most participants were neutral with regard to the statement. Nevertheless, the participants answered that question generally more positively.

To assess the AI literacy of our participants, we employed the ``Scale for the assessment of non-experts’ AI literacy'' \cite{laupichler_development_2023}. The participants responded to three key areas: Practical Application, Technical Understanding, and Critical Appraisal. On average, Practical Application scored a mean of 4.65 ($s = 0.66$), Technical Understanding received a mean score of 3.71 ($s = 0.46$), and Critical Appraisal had the highest mean score of 5.03 ($s = 0.20$). On average, our participants seem to be able to apply AI practically and approach it critically but do not entirely grasp all its technical aspects.

\begin{figure}
    \centering
    \begin{subfigure}[b]{0.49\textwidth}
        \centering
        \includegraphics[width=\textwidth]{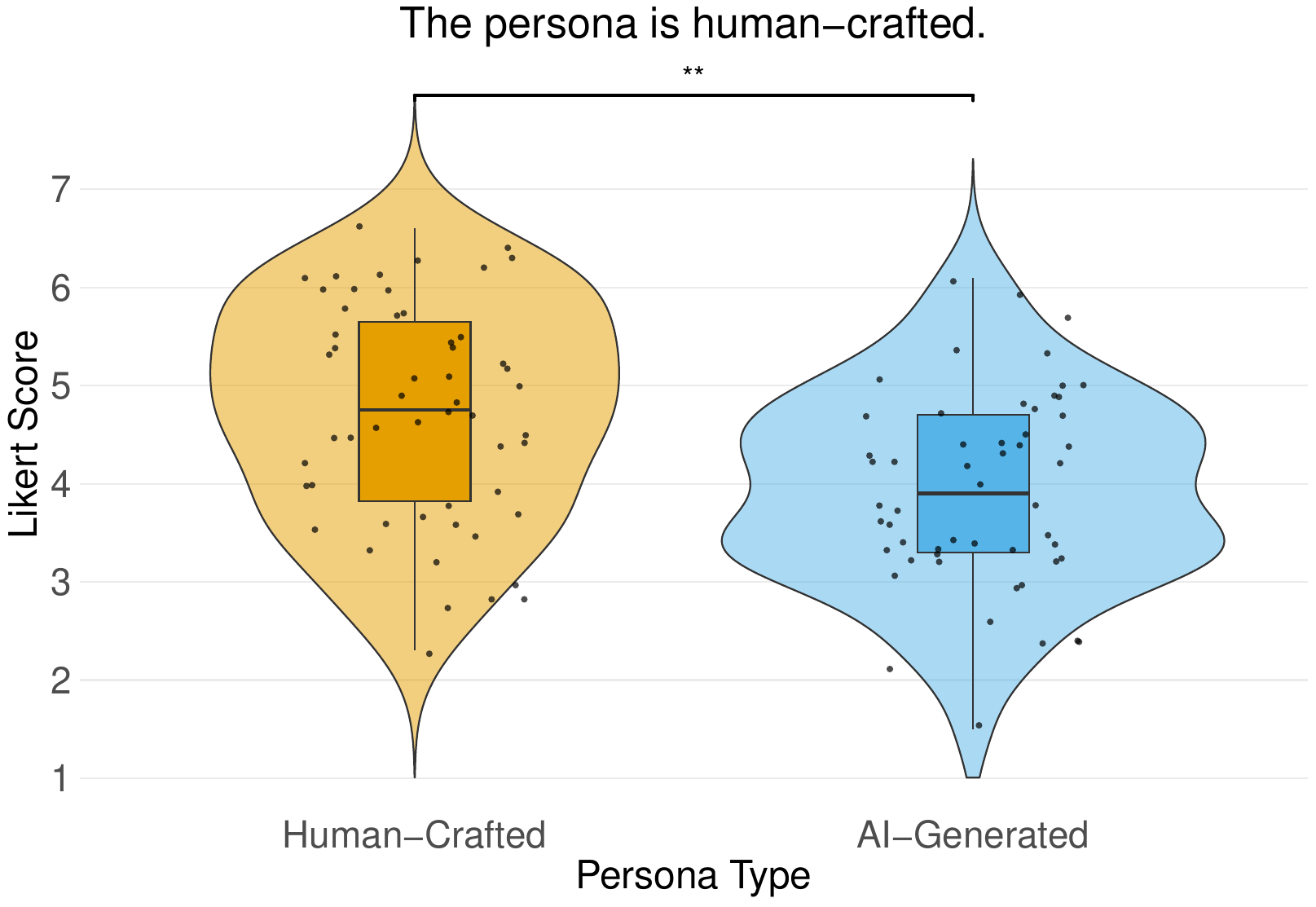}
        \caption{}
        \label{fig:s001}
    \end{subfigure}
    \hfill
    \begin{subfigure}[b]{0.49\textwidth}
        \centering
        \includegraphics[width=\textwidth]{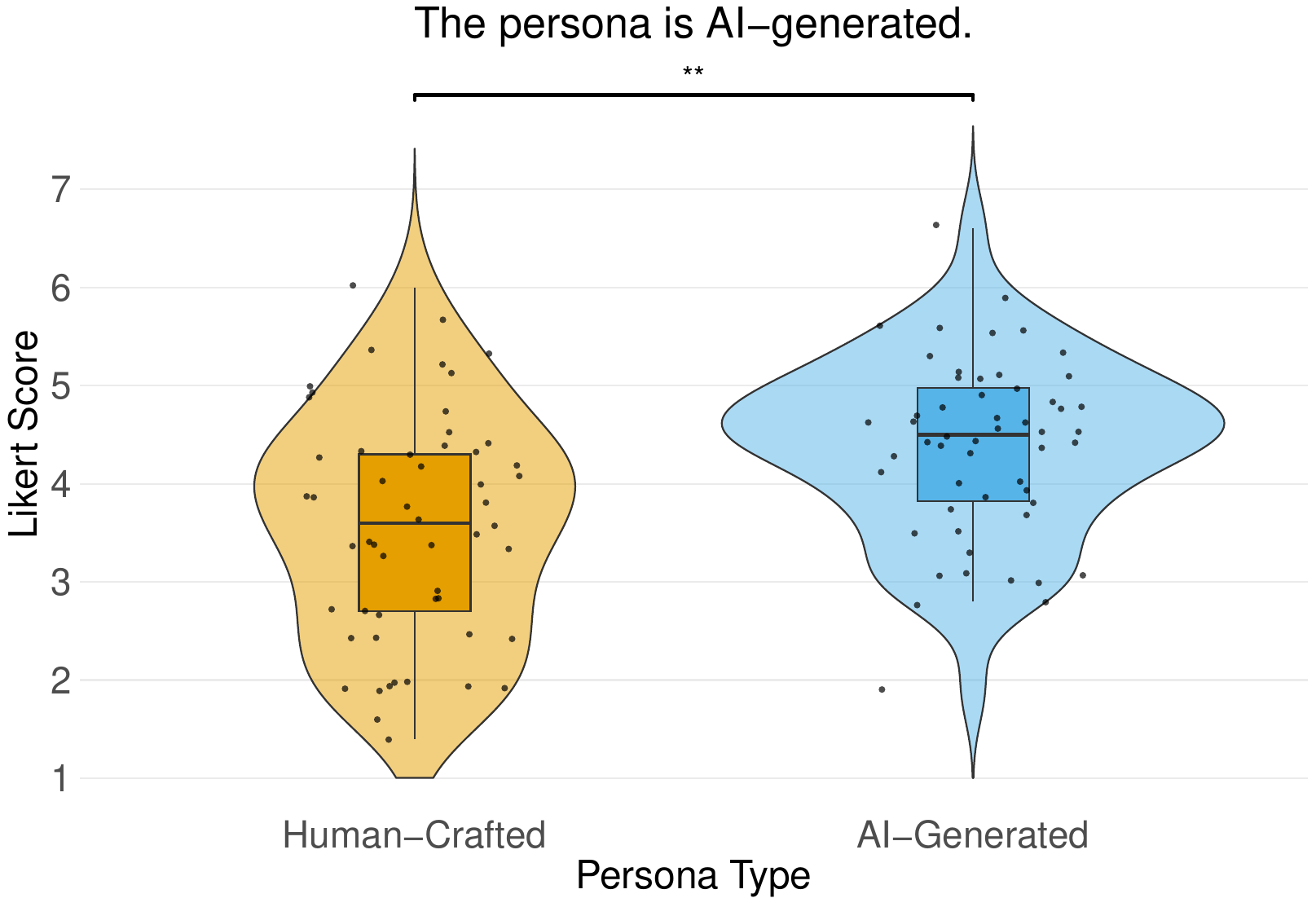}
        \caption{}
        \label{fig:s002}
    \end{subfigure}
    \caption{Violin plot comparing if participants could distinguish between human-crafted and AI-generated personas. \textbf{(a):} Participants could recognize human-crafted personas. \textbf{(b):} Participants could recognize AI-generated personas. Asterisk denote significant differences.}
    \Description{The Figure shows a Violin plot comparing if participants could distinguish between human-crafted and AI-generated personas. On one side, you can see that (a): Participants could recognize human-crafted personas with significance, and (b): Participants could recognize AI-generated personas with significance.}
    
    \label{fig:s001_s002}
\end{figure}

\subsection{Quantitative Results}
This section presents the findings of our study. We conducted Wilcoxon signed-rank tests to statistically compare the Likert ratings between human-designed and AI-generated personas. The statements that were ranked by our participants for each persona are shown in \autoref{table:aspects}. We averaged the Likert item ratings across personas for each participant to get a single value for human-crafted and generated personas, allowing for statistical comparison. We conducted a Wilcoxon signed-rank as a non-parametric test to compare the Likert ratings of the participants. The significance level was set at $\alpha = .05$. We also report the test statistics $V$, $Z$, and the effect size $r$\footnote{Generally, r = 0.1 is considered a small, r = 0.3 a medium, and r = 0.5 a large effect.}. We calculated the score of the constructs ``Clarity'' and ``Likability'' by calculating the mean score of the questions as suggested by the persona perception scale~\cite{SALMINEN2020102437}. 

\subsubsection{Distinguishing Between Human-Crafted and Generated Personas}
A Wilcoxon signed-rank test was performed to assess whether participants could distinguish between human personas and those generated by an LLM. Ratings were analyzed separately for human-crafted and AI-generated personas to enable direct comparisons. The results showed a significant difference between the two types of personas, with $V = 1088.5$, $Z = 2.98$, $p = .003$, $r = 0.405$. Human-crafted personas received higher Likert scores regarding the Likert statement that describes them as human-crafted, indicating that participants thought they were indeed human-crafted (see \autoref{fig:s001}). Consistently, we found a significant difference when participants encountered an AI-generated persona, with $V = 372.5$, $Z = -3.05$, $p = .002$, $r = -0.415$ (see \autoref{fig:s002}). Consequently, AI-generated personas scored higher for the corresponding Likert statement saying that they are AI-generated. 

\subsubsection{Informativeness for Design}
The Wilcoxon signed-rank test indicated a significant difference in the perceived informativeness between human-crafted and AI-generated personas, $V = 192$, $Z = -4.34$, $p < .001$, $r = -0.597$, with participants giving higher Likert scores to the generated personas when comparing them to the human-crafted ones (see \autoref{fig:s003}). 

\subsubsection{Believability}
\begin{figure}[h!]
    \centering
    \begin{subfigure}[b]{0.49\textwidth}
        \centering
        \includegraphics[width=\textwidth]{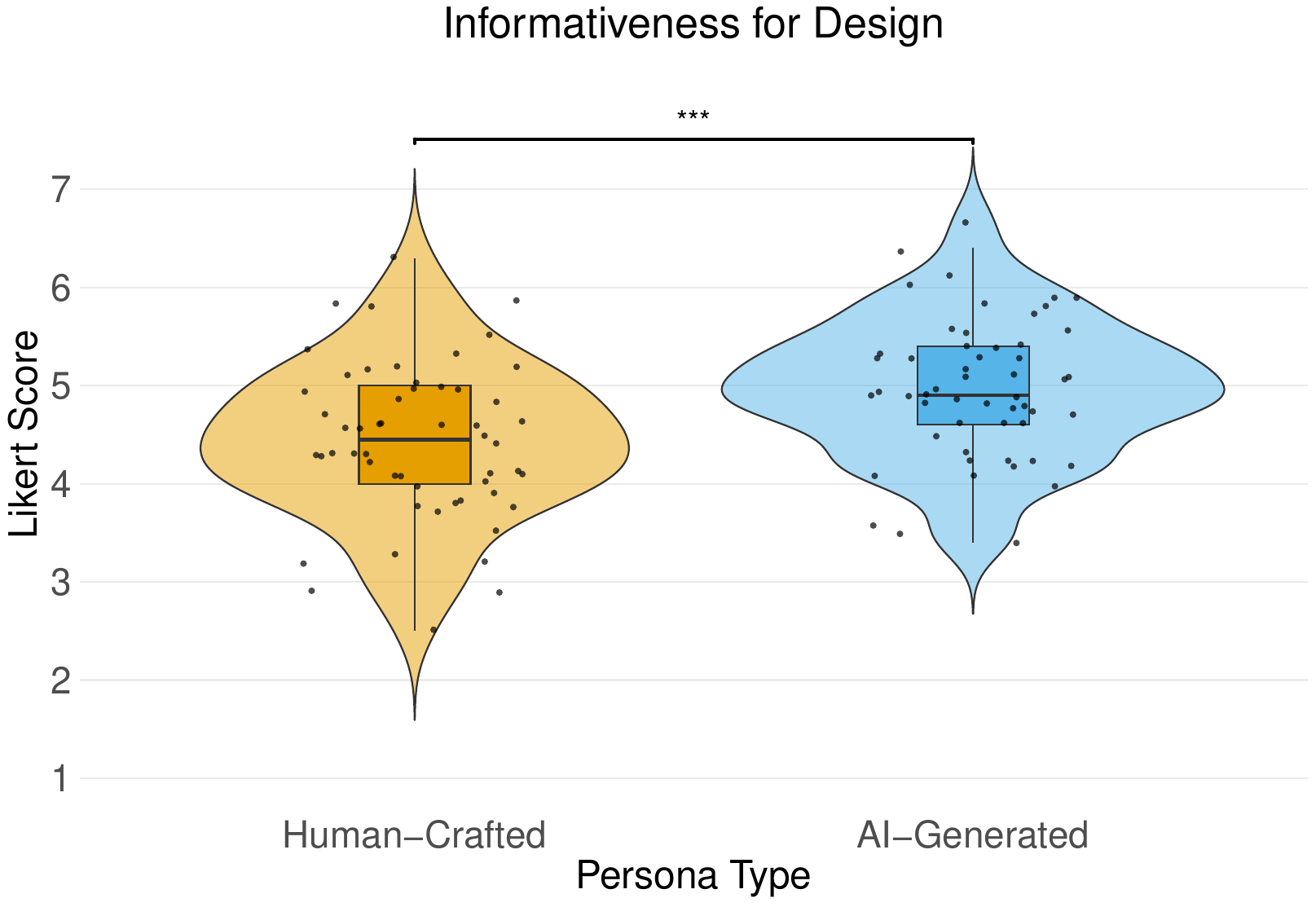}
        \caption{}
        \label{fig:s003}
    \end{subfigure}
    \hfill
    \begin{subfigure}[b]{0.49\textwidth}
        \centering
        \includegraphics[width=\textwidth]{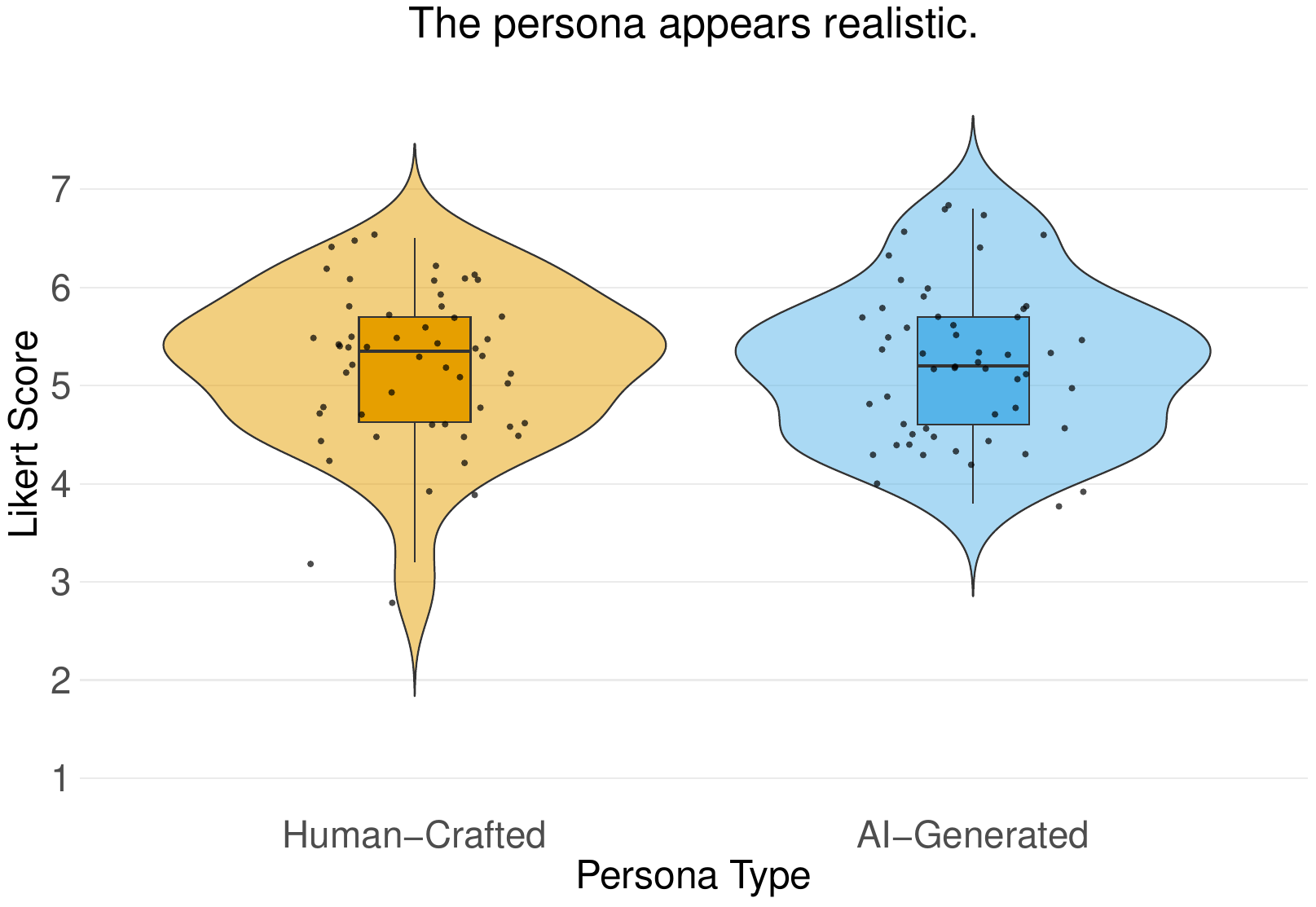}
        \caption{}
        \label{fig:s004}
    \end{subfigure}
    \caption{Violin plot comparing how participants rated the informativeness and the realism of human-crafted and AI-generated personas. \textbf{(a):} Participants rated the informativeness of AI-generated personas higher than that of human-crafted personas. We abbreviated the title of the figure with the persona aspect from \autoref{table:aspects} to increase the readability. \textbf{(b):} There was no significant difference between participants' ratings about the realism of AI-generated and human-crafted personas. Asterisk denote significant differences.}
    \Description{The Figure shows a Violin plot comparing how participants rated the informativeness and the realism of human-crafted and AI-generated personas. On one side, you can see that (a): Participants rated the informativeness of AI-generated personas significantly higher than of human-crafted personas, and (b): There was no significant difference between the ratings of participants about the realism of AI-generated and human-crafted personas.}
    \label{fig:s003_s004}
\end{figure}

The Wilcoxon signed-rank test showed no significant difference in the believability of human-crafted versus generated personas, $V = 667$, $Z = 0.05$, $p = .96$, $r = 0.006$ (see \autoref{fig:s004}). 

\subsubsection{Stereotypicality}
The Wilcoxon signed-rank test indicated a significant difference in the stereotypicality of human-crafted versus generated personas, $V = 356.5$, $Z = -2.24$, $p = .03$, $r = -0.305$. Generated personas received higher Likert scores than human-crafted ones (see \autoref{fig:s005}). 

\subsubsection{Positivity}
\begin{figure}
    \centering
    \begin{subfigure}[b]{0.49\textwidth}
        \centering
        \includegraphics[width=\textwidth]{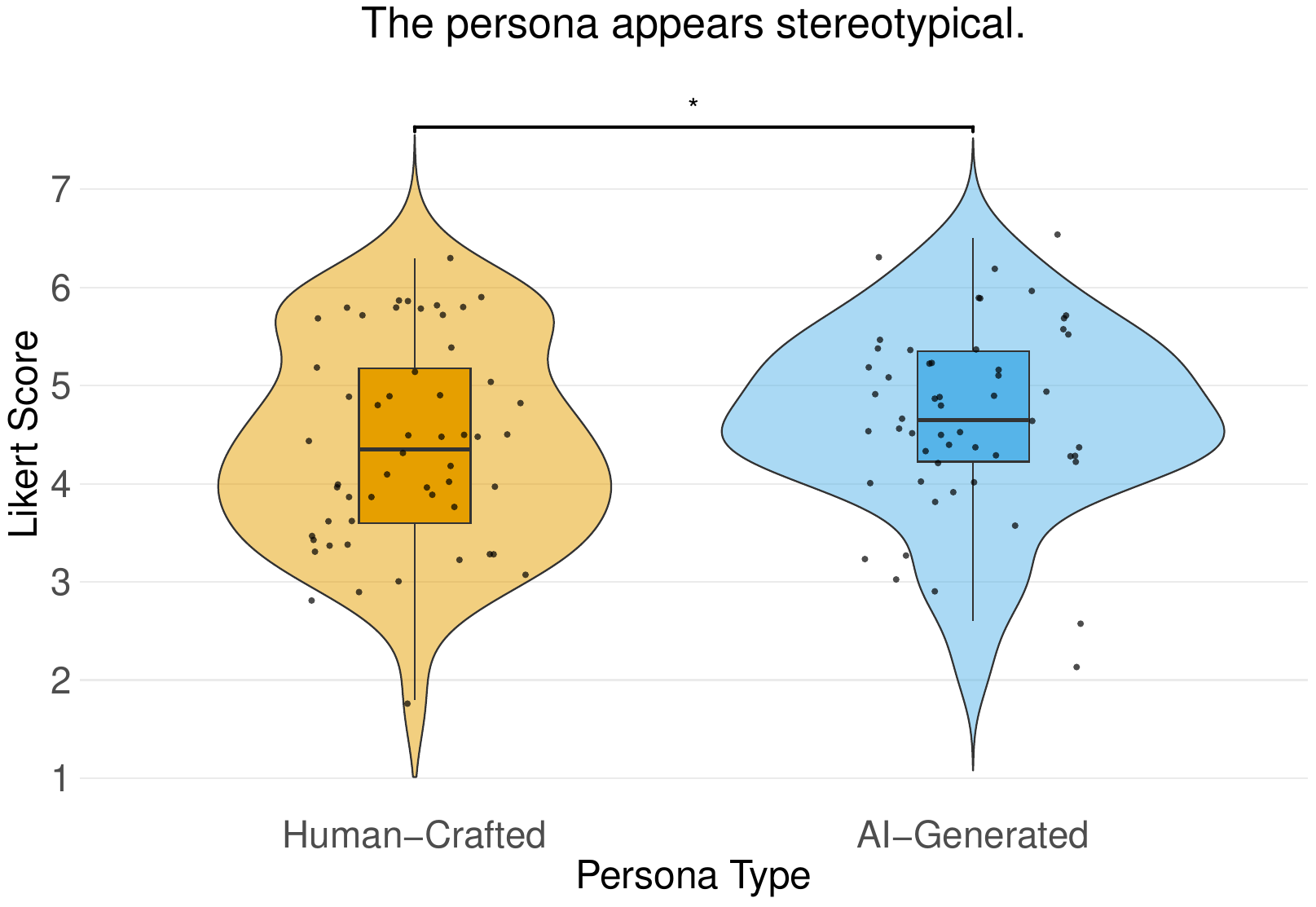}
        \caption{}
        \label{fig:s005}
    \end{subfigure}
    \hfill
    \begin{subfigure}[b]{0.49\textwidth}
        \centering
        \includegraphics[width=\textwidth]{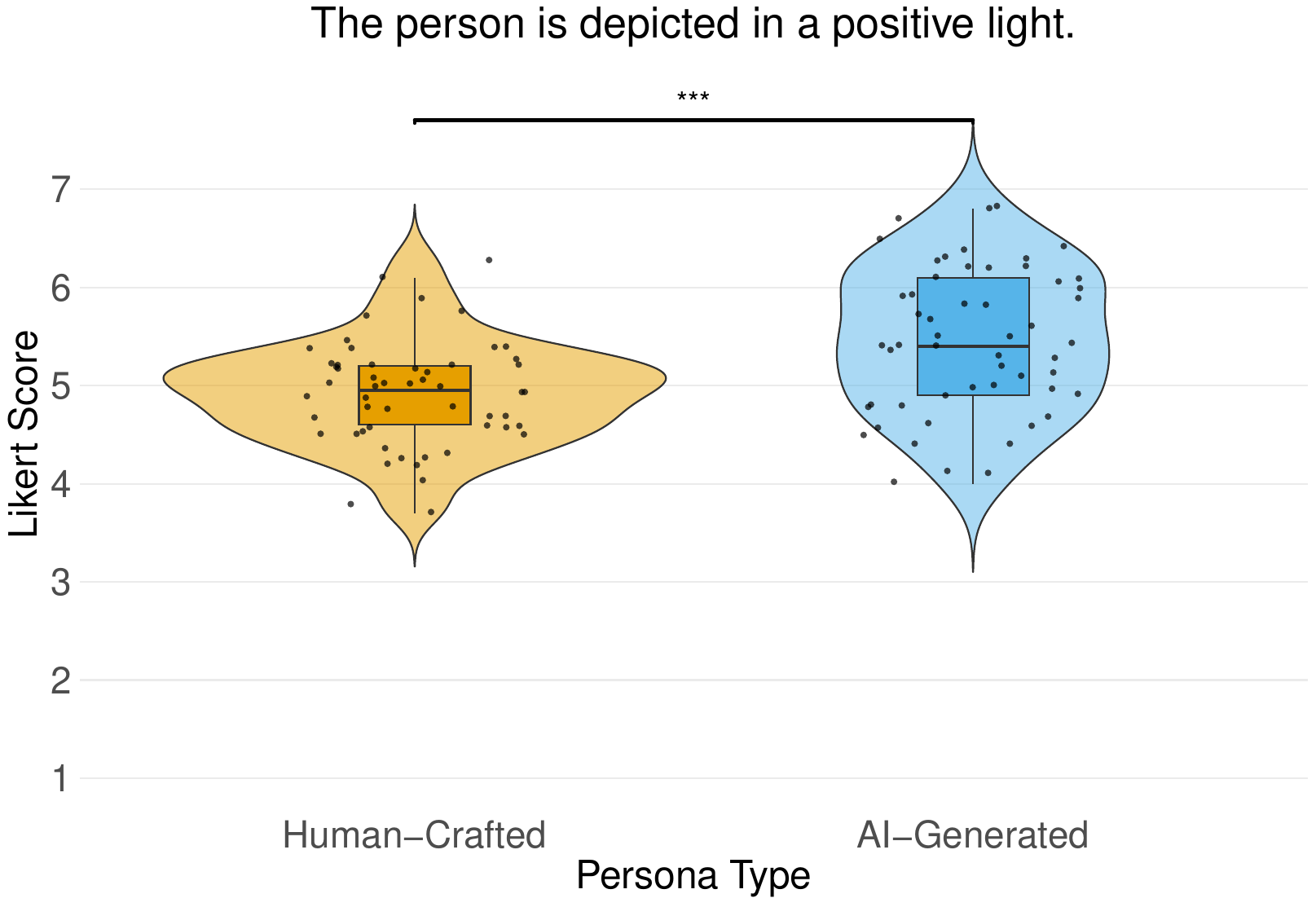}
        \caption{}
        \label{fig:s006}
    \end{subfigure}
    \caption{Violin plot comparing how participants rated the stereotypicality and the positivity of human-crafted and AI-generated personas. \textbf{(a):} Participants rated the stereotypicality of AI-generated personas higher than that of human-crafted personas. \textbf{(b):} Participants rated the positivity of AI-generated personas higher than of human-crafted personas. Asterisk denote significant differences.}
    \Description{The Figure shows a Violin plot comparing how participants rated the stereotypicality and the positivity of human-crafted and AI-generated personas. On one side, you can see that (a): Participants rated the stereotypicality of AI-generated personas significantly higher than of human-crafted personas, and (b): Participants rated the positivity of AI-generated personas higher than of human-crafted personas.}
    \label{fig:s005_s006}
\end{figure}

The Wilcoxon signed-rank test also revealed a significant difference in the positivity of human-crafted and generated personas, $V = 203.5$, $Z = -4.31$, $p < .001$, $r = -0.586$. Likert scores were higher for AI-generated personas than human-crafted ones (see \autoref{fig:s006}). 

\subsubsection{Relatability}
The Wilcoxon signed-rank test did not indicate a significant difference in the believability between human-crafted and AI-generated personas, $V = 593.5$, $Z = -0.16$, $p = .85$, $r = -0.025$ (see \autoref{fig:s007}). 

\subsubsection{Consistency}
\begin{figure}[h!]
    \centering
    \begin{subfigure}[b]{0.49\textwidth}
        \centering
        \includegraphics[width=\textwidth]{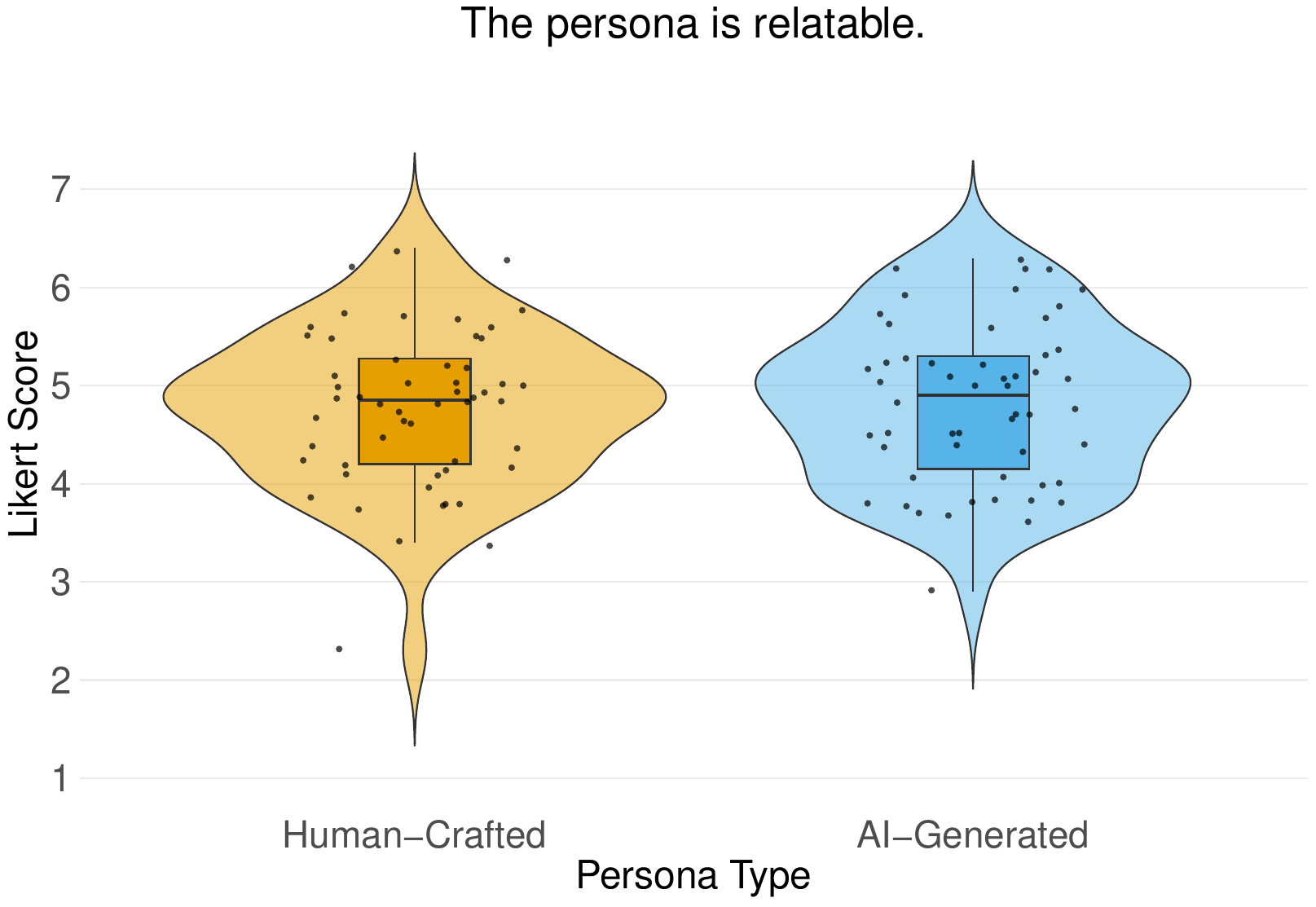}
        \caption{}
        \label{fig:s007}
    \end{subfigure}
    \hfill
    \begin{subfigure}[b]{0.49\textwidth}
        \centering
        \includegraphics[width=\textwidth]{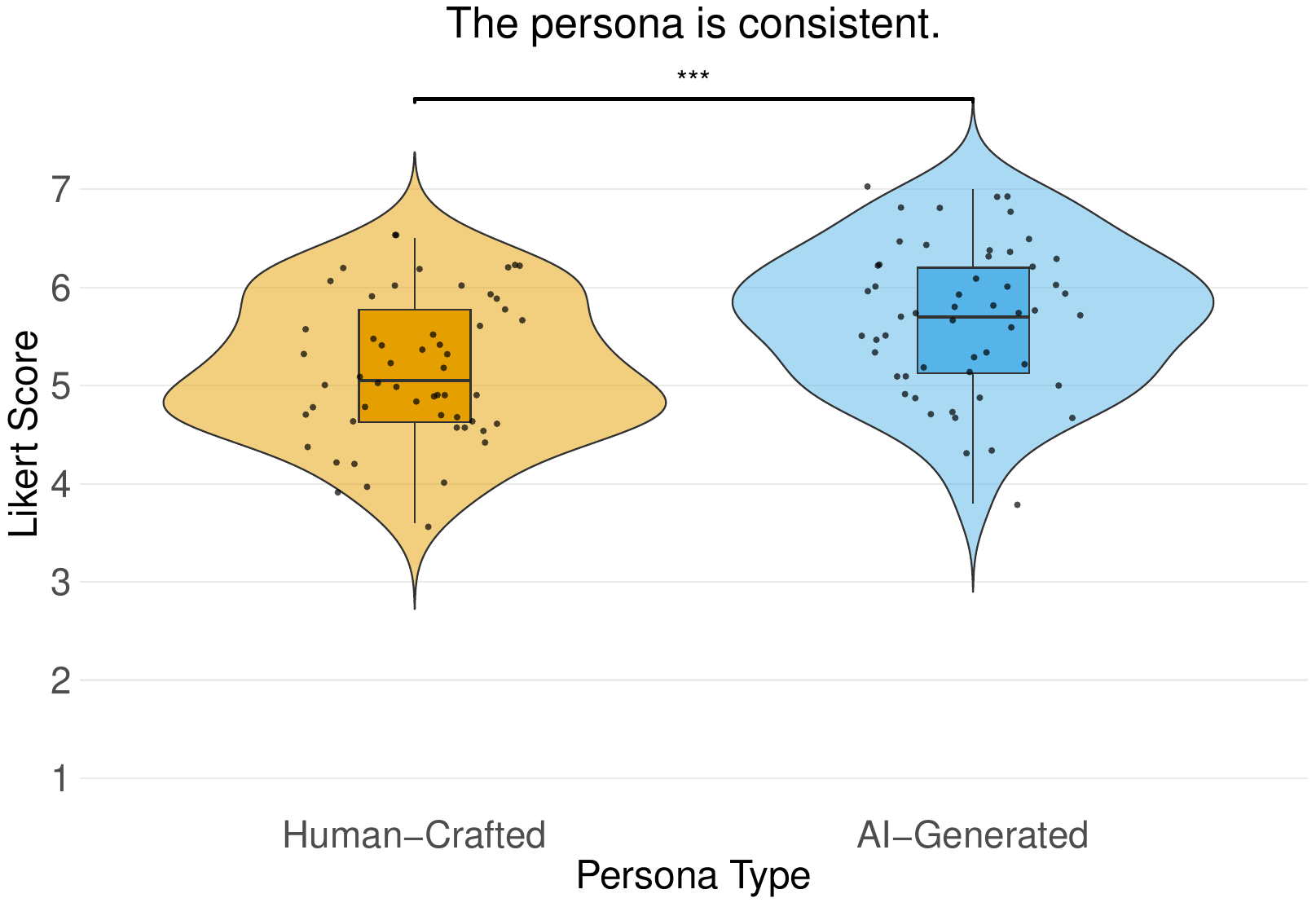}
        \caption{}
        \label{fig:s008}
    \end{subfigure}
    \caption{Violin plot comparing how participants rated the relatability and the consistency of human-crafted and AI-generated personas. \textbf{(a):} Participants rated the relatability of AI-generated personas higher than that of human-crafted personas. \textbf{(b):} Participants rated the consistency of AI-generated personas higher than of human-crafted personas. Asterisk denote significant differences.}
    \Description{The Figure shows a Violin plot comparing how participants rated the relatability and the consistency of human-crafted and AI-generated personas. On one side, you can see that (a): Participants rated the relatability of AI-generated personas significantly higher than of human-crafted personas, and (b): Participants rated the consistency of AI-generated personas significantly higher than of human-crafted personas.}
    \label{fig:s007_s008}
\end{figure}

The Wilcoxon signed-rank test showed a significant difference in the consistency of human-crafted versus AI-generated personas, $V = 155.5$, $Z = -4.40$, $p < .001$, $r = -0.599$, with participants rating the AI-generated personas higher in consistency (see \autoref{fig:s008}). 

\subsubsection{Clarity}
The Wilcoxon signed-rank test showed a significant difference in the clarity between human-crafted and generated personas, $V = 223.5$, $Z = -4.08$, $p < .001$, $r = -0.556$, with higher Likert scores for AI-generated personas (see \autoref{fig:clarity}). 

\subsubsection{Likability}
\begin{figure}[h!]
    \centering
    \begin{subfigure}[b]{0.49\textwidth}
        \centering
        \includegraphics[width=\textwidth]{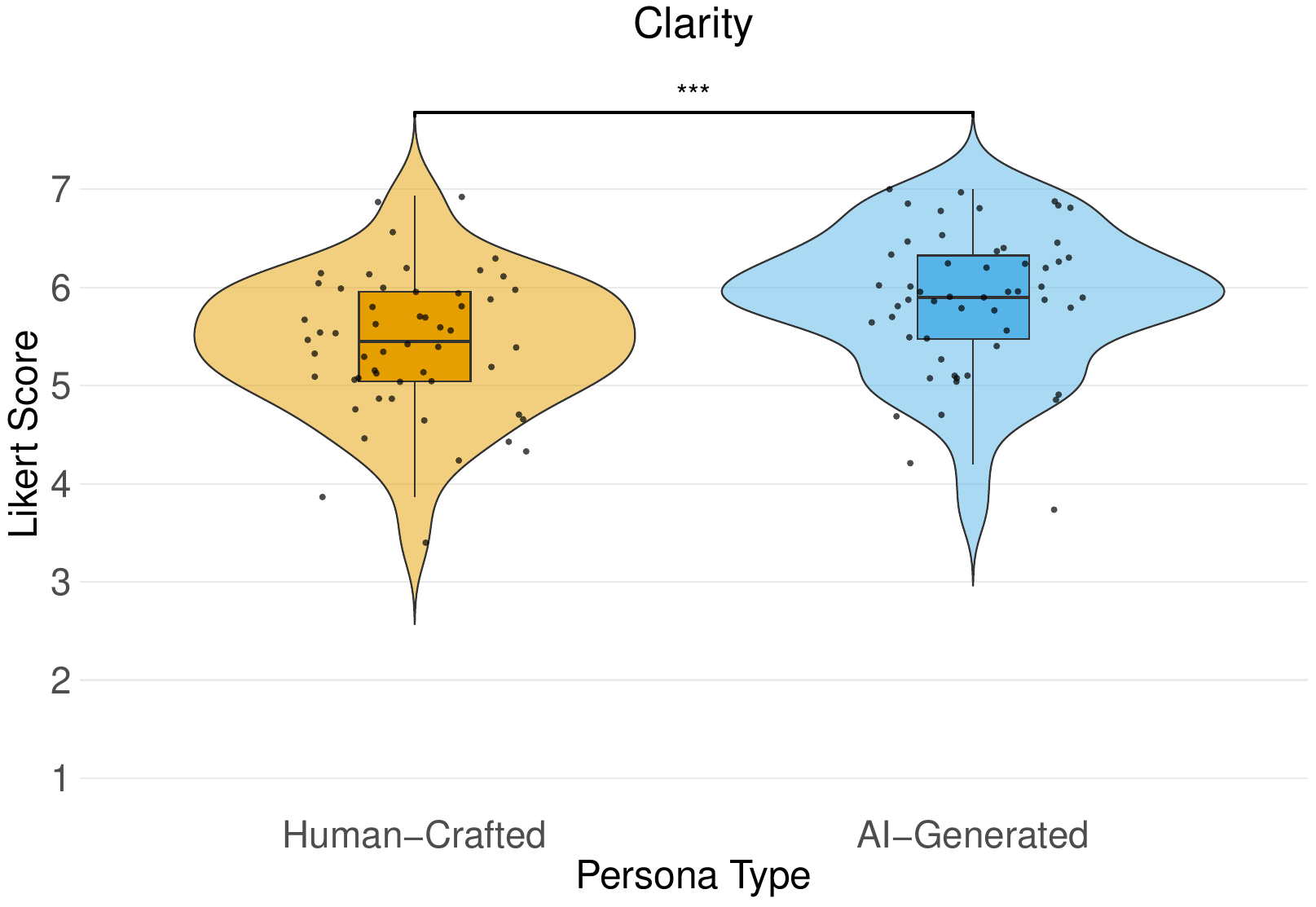}
        \caption{}
        \label{fig:clarity}
    \end{subfigure}
    \hfill
    \begin{subfigure}[b]{0.49\textwidth}
        \centering
        \includegraphics[width=\textwidth]{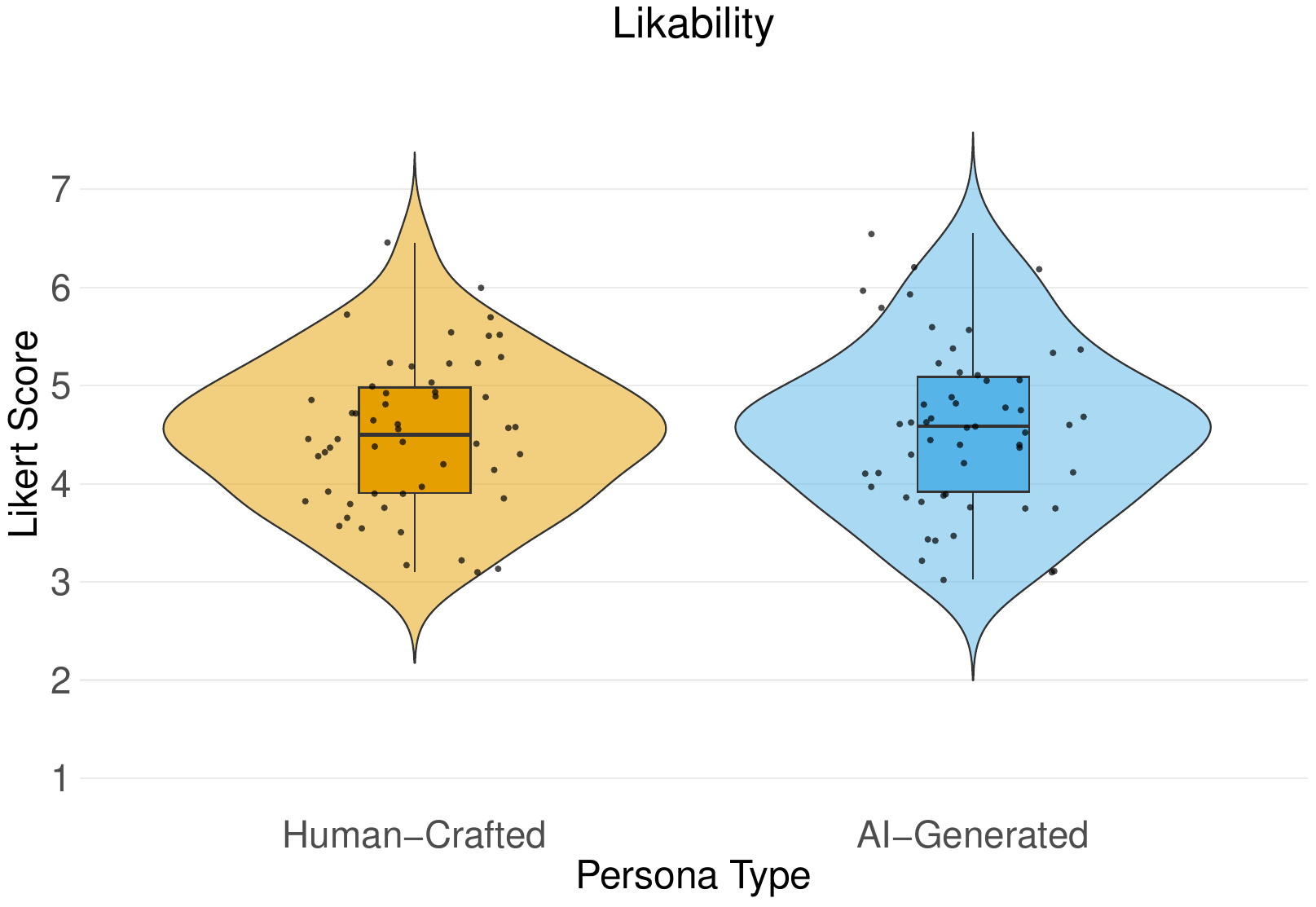}
        \caption{}
        \label{fig:likability}
    \end{subfigure}
    \caption{Violin plot comparing how participants rated the clarity and the likability of human-crafted and AI-generated personas. \textbf{(a):} Participants rated the clarity of AI-generated personas higher than that of human-crafted personas. \textbf{(b):} There was no significant difference between the ratings of participants about the likability of AI-generated and human-crafted personas. Asterisk denote significant differences.}
    \Description{The Figure shows a Violin plot comparing how participants rated the clarity and the likability of human-crafted and AI-generated personas. On one side, you can see that (a): Participants rated the clarity of AI-generated personas significantly higher than of human-crafted personas, and (b): There was no significant difference between the ratings of participants about the likability of AI-generated and human-crafted personas.}
    \label{fig:clarity_likability}
\end{figure}

The Wilcoxon signed-rank test indicated no significant difference in likability between human-crafted and generated personas, $V = 672$, $Z = -0.38$, $p = .71$, $r = -0.052$ (see \autoref{fig:likability}). 

\subsection{Qualitative Results}
We started the qualitative analysis by categorizing the free-text responses based on their classification as human-crafted, AI-generated, neutral, or conflicted. Then, we applied a second tag to indicate the correctness of the participant's classification. For instance, if a participant labeled a persona as human-crafted when it was AI-generated, we marked the classification as incorrect. Conversely, a tag indicating a correct choice was applied when the participant correctly identified the origin of a persona. Neutral classifications were assigned if participants rated both ``The persona is human-crafted'' and ``The persona is AI-generated'' with a score of four. A conflicted tag was given when both questions were rated below or above four. This tagging system allowed us to code the free-text answers about the accuracy of the participants' choices.

Two authors coded the free-text responses. We randomly assigned 25\% of the written responses from when participants encountered human-crafted and AI-generated content. Following Blandford et al.~\cite{blandford2016qualitative}, we applied an inductive coding approach to generate an initial set of codes and construct coding trees. Afterward, we conducted a code adjustment session where we reviewed the initial coding tree and distributed the remaining free-text responses among three coders. Once the entire data set was coded, we engaged in iterative discussions using axial coding to refine the final coding tree and identify recurring patterns. We extracted the five themes \emph{Writing}, \emph{Information}, \emph{Stereotypicality}, \emph{Realism and Appeal}, and \emph{Positive and Negative Appeal} from our coding process, which captured participants' perceptions of personas as either real, artificial, or undecided. Given the exploratory nature of our study, we determined that an open-ended thematic analysis, as outlined by Blandford et al.~\cite{blandford2016qualitative}, was the most appropriate approach. This analysis method, rooted in interpretivism, is commonly employed in HCI research.

\subsubsection{Writing}
In the evaluation of personas generated by LLMs, one of the recurring themes identified in participants' responses was the distinctive writing style of the personas. Participants often pointed to specific elements such as sentence structure, grammatical errors, and vocabulary choice when forming their impressions. These writing characteristics played a key role in shaping how the participants perceived the personas, often serving as subtle indicators of whether the text was likely human-generated or AI-generated:

\begin{quote}
    \textit{``When compared to the previous personas that are believed to be AI generated, there is a striking resemblance in the words and phrases used, such as "deep-seated", "Outside of her professional life,", and "global understanding".  This is no coincidence; I strongly believe these are all connected due to the work of Artificial Intelligence.''} (P15, AI-generated persona correctly rated)
\end{quote} 


While many participants cited the writing style as a key reason for their rating of the personas, their interpretations of specific elements, such as grammatical errors, varied. Some participants argued that the presence of grammatical mistakes indicated that the text was likely human-generated, as they believed an AI would not produce such errors:

\begin{quote}
    \textit{``Again found some grammar issues within the text AI wouldn’t do this.''} (P18, human-crafted persona correctly rated)
\end{quote} 

On the other hand, others suggested the opposite, claiming that AI and LLMs do not generate grammatically flawless text, thus implying that errors were more characteristic of AI writing. These contrasting views highlight the complexity of how participants perceived AI-generated content, with grammatical accuracy being a point of contention in determining whether the writing felt more human- or machine-like. Consequently, other participants were misguided by grammatical errors:

\begin{quote}
    \textit{``way too many errors in the description give away that it is not human generated''} (P29, human-crafted persona falsely rated)
\end{quote} 

Overall, participants identified the language used in the personas as a common factor influencing their ratings. There was a consensus among participants that AI-generated personas often adopted a "robotic tone," which stood out in contrast to more natural human speech. The vocabulary used by these personas was frequently described as "unusual" or somewhat disconnected from everyday language, further reinforcing the perception that the text lacked human-like fluidity and warmth. This shared impression of the AI personas' language as overly formal, rigid, or unnatural contributed to a sense of distance between the personas and the participants, shaping their overall assessment of the personas' authenticity and reliability:

\begin{quote}
    \textit{``I believe this was AI-generated due to the use of some words such as "honed", "permeating" and "further cementing" that are not commonly used in conversations.''} (P22, AI-generated persona correctly rated)
\end{quote} 

Furthermore, participants noted that human-crafted personas had a more conversational and relatable tone, resembling how a person would naturally speak to someone else. They emphasized that these personas felt more personal and engaging, with language that flowed in a way that mirrored typical human interactions. This conversational style made the human-crafted personas appear more authentic, as if they genuinely addressed the reader. In contrast to the AI-generated personas, often described as robotic or detached, human-crafted personas were seen as more capable of forming a connection through their more intuitive and familiar language:

\begin{quote}
    \textit{``Definitely human written, no doubt in my mind. The use of language here reads like they're speaking to someone while they're writing it down- more of an inner monologue, or like someone was describing someone they knew to someone else. [...]''} (P35 human-crafted persona correctly rated)
\end{quote} 

The participants noticed repetition in the persona descriptions. Surprisingly, repetitions led to both correct and wrong ratings. The participants associated the repetition with AI generation, especially if the persona's name was repeated:

\begin{quote}
    \textit{``The descriptions seem to have very repetitive words which makes me suspect it's AI-generated.''} (P41, AI-generated persona correctly rated)
\end{quote} 

However, participants also rated human-crafted personas as AI-generated because of repetitive writing:

\begin{quote}
    \textit{``The arrangement of the write up and the continual use of the person's name influenced me to decide it is an AI-generated persona.''} (P42, human-crafted persona falsely rated)
\end{quote} 

\subsubsection{Information}
Another common theme in participants' responses was the nature of the information provided in the persona descriptions. Participants often associated AI-generated personas with an abundance of details that, while extensive, felt less personal and occasionally irrelevant. They remarked that AI personas tended to include information that seemed extraneous or overly factual, lacking the nuanced, human touch that makes descriptions feel personalized. This tendency to over-provide impersonal or unnecessary details further contributed to the perception that the AI-generated personas were less relatable and more mechanical, reinforcing the idea that they lacked the selectiveness found in human-crafted descriptions:

\begin{quote}
    \textit{``This persona feels AI-generated because there are many unnecessary details, which AI is kind of notorious for doing at least in my experience''} (P21, AI-generated persona, correctly rated)
\end{quote} 

\begin{quote}
    \textit{``This persona seems human because there aren't unnecessary details and everything mentioned is realistic and feasible''} (P21, AI-generated persona falsely rated)
\end{quote} 

Whenever participants observed that a persona description included emotional or personal details, they were more likely to rate the personas as human-crafted. These emotional elements, such as expressions of feelings, experiences, or personal anecdotes, made the personas feel more genuine and relatable. Participants associated these details with the complexity and depth of human communication, which often involves subtle emotional cues. The presence of such personal information helped to distinguish human-crafted personas from AI-generated ones, which were typically viewed as more neutral or detached. This connection to emotional content reinforced the belief that the persona was created with a human touch, leading to higher ratings in authenticity and engagement:

\begin{quote}
    \textit{``Detailed, consistent, deliberate balance between included work and personal details. Some minor grammatical errors.''} (P36, human-crafted persona correctly rated)
\end{quote}

When AI-generated personas included more personal or emotional details, participants were often misled into believing they were human-crafted. These personal touches, such as references to experiences, emotions, or unique characteristics, made the personas appear more authentic and relatable, blurring the line between AI and human-generated content. By incorporating these human-like elements, the AI personas effectively mimicked the complexity and depth that participants typically associated with human authorship. As a result, the participants were more likely to overlook the artificial nature of the personas, rating them as if they had been crafted by a human rather than generated by an algorithm. This demonstrates how including personal details can significantly enhance the perceived authenticity of AI personas:

\begin{quote}
    \textit{``Nancy Jackson's persona appears human-crafted due to the detailed and nuanced depiction of her professional background, motivations, and personal interests. The inclusion of her passion for cultural diversity and traveling reflects a level of personalization and depth that suggests a human creator. AI-generated personas often lack such specific and individualized details, making the nuanced elements of Nancy’s profile indicative of human authorship.''} (P39, AI-generated persona falsely rated)
\end{quote}

Some AI-generated personas included free-time activities that appeared directly related to the persona's occupation, which participants often recognized as a clear indicator of AI generation. These activities, while logically consistent with the persona's professional background, lacked the diversity and spontaneity typically seen in human-crafted personas. Participants noted that real people tend to have hobbies and interests that may not always align with their work, adding to their individuality. The overly predictable nature of AI-generated personas' free-time activities made them feel less authentic and more formulaic, leading participants to rate them as AI-generated rather than human-crafted. This highlighted how rigid connections between work and personal life can diminish the perceived realism of AI personas:

\begin{quote}
    \textit{``His hobby is reading biographies of successful business leaders? He doesn't sound real at all. It's all generic and lacks detail. Probably AI.''} (P19, AI-generated persona correctly rated)
\end{quote}

Participants of our study noticed that some personas have an attitude to technology that appears to be rather critical instead of open and positive. This was reported to be a reason for rating personas as human-crafted.

\begin{quote}
    \textit{``The fact it mentions him being worried about AI has me thinking it's more likely to be human-crafted because I just couldn't imagine the AI including that in its description of someone.''} (P40, human-crafted persona correctly rated) 
\end{quote}

In addition to the presence and details included in a persona description, participants also argued about the consistency of a persona's information. Most of the participants argued that consistent personas are human-crafted:

\begin{quote}
    \textit{``Anne not accepting the use of new technology like the smartphone or work tools is consistent with people of older age.''} (P23, human-crafted persona correctly rated)
\end{quote}

Consistent AI-generated personas were also rated as human-crafted because of their consistency:

\begin{quote}
    \textit{``The features so described are consistent throughout the narrative''} (P23, AI-generated persona falsely rated)
\end{quote}

\subsubsection{Stereotypicality}
Participants frequently remarked that many personas appeared stereotypical or generic, a theme that emerged for AI-generated and human-crafted personas. However, their interpretations of this stereotypicality influenced their evaluations differently. Some participants noted that when a persona was overly stereotypical or lacked unique traits, they were more inclined to believe it was AI-generated. They argued that AI tends to rely on familiar patterns, making it more likely to create personas that fit into generic molds without the nuance or individuality of real people. For these participants, the lack of depth or complexity in such personas indicated that they were machine-generated, as human-crafted personas were expected to show more diversity and distinctiveness in their characteristics:

\begin{quote}
    \textit{``I rated as AI-generated because this persona feels quite stereotypical. A 20 something IT guy that builds PC and games? Very common in real life, but also a very common stereotype that AI could easily generate.''} (P16, AI-generated persona correctly rated)   
\end{quote}

On the other hand, participants argued that less stereotypical personas were more likely to be human-crafted. These personas, characterized by unique or unconventional traits, felt more personalized and authentic, leading participants to believe humans designed them. The distinctiveness and originality of the non-stereotypical personas made them appear more reflective of real human complexity, as they defied the generic patterns often associated with AI-generated content. For these participants, the nuanced and individualized nature of such personas was a strong indicator of human authorship, reinforcing the belief that humans are better at creating diverse and less predictable personas:

\begin{quote}
    \textit{``The persona of Chris feels unique and non-stereotypical with characteristics such as his love of birds which is not exactly usual for a 10 year old boy in America. I feel like an AI-generated persona might stick to more stereotypical interests.''} (P16, human-crafted persona correctly rated)   
\end{quote} 

The stereotypical nature of some human-crafted personas even misleads participants into rating them as AI-generated, especially if the participants perceive the persona as believable or realistic:

\begin{quote}
    \textit{``The story does feel believable as a human but it seems too stereotypical to be human crafted''} (P54, human-crafted persona falsely rated)   
\end{quote} 

\subsubsection{Realism and Appeal}
In analyzing the participants' written responses, a common theme emerged: the perception that specific personas appeared realistic, believable, and lifelike. Participants frequently used these qualities as key indicators when arguing that a persona was human-crafted. They emphasized that the more a persona felt authentic and mirrored real human behavior, thoughts, and emotions, the more likely they attribute its creation to a human rather than an AI. The lifelike and relatable nature of these personas, with their believable characteristics and interactions, made participants feel that only a human could craft such a nuanced and convincing depiction, further reinforcing their judgments of authenticity:

\begin{quote}
    \textit{``Oh dear! I know he's not a real person but wow do I feel sorry for Peter, and am hoping he can move forward and improve his life. As I had quite an emotional first reaction to this persona, I'm going to rate it as human-crafted. He feels like he could exist in real life. Not necessarily as a good person, or likable, but real.''} (P16, human-crafted persona correctly rated) 
\end{quote}

Accordingly, a realistic depiction also misled participants in the case of AI-generated personas. When AI-generated personas exhibited believable and lifelike traits, participants were more likely to mistake them for being human-crafted. The realistic portrayal, including natural dialogue, relatable behaviors, and plausible personal details, blurred the line between AI and human authorship. This sense of authenticity created by the AI misled participants into thinking the persona was the product of human creativity, highlighting how effective AI can be at mimicking the complexity of human expression when it incorporates realistic elements:

\begin{quote}
    \textit{``The details influenced my decision to rate the persona as human-crafted and the information seemed consistent and realistic across every aspect.''} (P27, AI-generated persona falsely rated) 
\end{quote}

Participants partly associated a stereotypical description with personas that reflect reality:

\begin{quote}
    \textit{``This could have been human-crafted based on a real person. Stereotypical description of someone from this age group that liked to garden and is afraid of technology.''} (P22, human-crafted persona correctly rated) 
\end{quote}

This association was also evident in participants' uncertainty when rating non-stereotypical personas, as they often appeared unrealistic. While participants generally associated non-stereotypical personas with human authorship, the lack of familiar or conventional traits sometimes made these personas feel less believable. As a result, participants found it challenging to determine whether such personas were AI-generated or human-crafted confidently. The unusual or unexpected characteristics of these personas created ambiguity, making participants question their initial assumptions about what constitutes a realistic, human-like portrayal:

\begin{quote}
    \textit{``It feels a bit made up. Nobody plays Pokemon Go anymore, not since like 2016. On the other hand, it's believeable and there is some specific detail. I'm not sure about this one.''} (P19, human-crafted persona neutrally rated) 
\end{quote}

Furthermore, participants argued that ``likeable'' personas are human-crafted:

\begin{quote}
    \textit{``This persona is human crafted. he is interesting and likable''} (P38, human-crafted persona correctly rated) 
\end{quote}

\subsubsection{Positive and Negative Descriptions}
A common reason participants rated personas as AI-generated was the presence of an overly positive description. Many participants remarked that some AI-generated personas seemed "too positive," presenting an unrealistic or excessively idealized depiction. This overly optimistic tone, where the persona exhibited no apparent flaws or challenges, was perceived as lacking the nuance and complexity of real human behavior. Participants viewed this as a hallmark of AI generation, as they expected human-crafted personas to reflect a more balanced and authentic representation of strengths and weaknesses. The excessive positivity made the personas feel artificial, leading to lower ratings in terms of realism and authenticity:

\begin{quote}
    \textit{``[...] There also doesn't appear to be any inherently negative things said, even just a small thing- like with the previous example of the gardener with a back problem. This profile just seems too perfect.''} (P35, AI-generated persona correctly rated) 
\end{quote}

Furthermore, participants were more likely to rate personas as human-crafted if depicted in a more negative or balanced light. Personas that included imperfections, flaws, or challenges were perceived as more authentic and realistic, leading participants to associate them with human authorship. Including negative traits or struggles made the personas feel more relatable and complex, as these characteristics better reflected the nuanced nature of real human experiences. Participants saw this more grounded and less idealized portrayal as something a human would be more likely to create, unlike the overly positive and polished personas often associated with AI generation:

\begin{quote}
    \textit{``seems like a very basic description of a normal person, I feel like ai would not be so harsh''} (P33, human-crafted persona correctly rated) 
\end{quote}
\section{Discussion}
We conducted a study to investigate differences between human-crafted and AI-generated personas. In the previous sections, we reported the results of our persona crafting and generation process and our online survey's quantitative and qualitative results. In the following section, we discuss the results of our overall research questions.

\subsection{Distinguishing between Human-Crafted and AI-Generated Personas}
Our first research question aimed to investigate to what extent participants can distinguish between human-crafted and AI-generated personas. 
As shown in our quantitative results, we found a significant difference between the rating of human-crafted and AI-generated personas. Human-crafted personas achieved higher Likert scale ratings when participants were asked if the shown persona was human-crafted. Accordingly, in the case of AI-generated personas, participants rated the statement that a persona is AI-generated significantly higher on a Likert scale. Consequently, our results do not replicate the work by Schuller et al. \cite{schuller_generating_2024}, who reported that AI-generated personas were indistinguishable from human-crafted personas, suggesting similar quality and acceptance. 
Schuller et al. employed a non-parametric Wilcoxon rank-sum test on results from 11 participants, which did not reject the null hypothesis. While in their case, the null hypothesis assumes that participants could not distinguish between AI-generated and human-crafted personas, not rejecting the null hypothesis does not prove it. A significant effect can be absent due to the low sample size. 
By asking a group of 54 participants and focusing on novice users, we tried to investigate the topic further, more specifically. These factors may have led to a significant difference in the measured results.
\textbf{Thus, we answer RQ1:}

\begin{flushleft}   
\setlength{\fboxsep}{10pt}
\setlength{\fboxrule}{2pt}
\fcolorbox{gray!60}{white}{%
    \parbox{.89\columnwidth}{%
    \textbf{RQ1:}  Our results indicate that participants could distinguish between the human-crafted and AI-generated personas. When asked whether a persona was human-crafted, participants rated human-crafted personas significantly higher on a 7-point Likert scale ($p = .003$). Similarly, when asked whether a persona was AI-generated, participants rated AI-generated personas significantly higher ($p = .002$). These findings suggest that participants can identify differences between personas based on their origin, demonstrating a perceptual distinction between human-crafted and AI-generated personas.
    }
}
\end{flushleft}

\subsection{Perception of Human-Crafted and AI-Generated Personas}
Our second research question aimed to investigate which features in a persona's description affect how this persona is perceived and why participants rate a persona as human-crafted or AI-generated. Therefore, we collected ratings for different aspects of personas based on related literature. Our quantitative results replicated the work by Salminen et al. \cite{salminen_deus_2024}. We found significant differences between human-crafted and AI-generated personas regarding their informativeness, positivity, and consistency. Participants rated these aspects significantly higher on a 7-point Likert scale for AI-generated persona. Additionally, we found a significant difference in the stereotypicality of persona description. Participants rated AI-generated personas as significantly more stereotypical when answering a 7-point Likert scale question. Related work stated that AI-generated personas were perceived as non-stereotypical \cite{salminen_deus_2024}.  Furthermore, we found a significant difference regarding the clarity of personas. Participants rated AI-generated personas significantly higher than human-crafted personas.  We could not find any significant differences in realism, relatability, or likability between human-crafted and AI-generated personas. \textbf{We answer RQ2.1:}

\begin{flushleft}   
\setlength{\fboxsep}{10pt}
\setlength{\fboxrule}{2pt}
\fcolorbox{gray!60}{white}{%
    \parbox{.89\columnwidth}{%
    \textbf{RQ2.1:} Participants rated AI-generated personas as significantly clearer, more consistent, more positive, and more informative than the human-crafted personas. Despite these strengths, AI-generated personas were also perceived as significantly more stereotypical. This suggests that AI-generated personas may excel in certain areas, such as clarity and consistency, but they also have a higher risk of reinforcing stereotypes. 
    }
}
\end{flushleft}

\subsection{Features That Distinguish Human-Crafted and AI-Generated Personas}
Our qualitative results showed that the main difference between human-crafted and AI-generated personas was the writing and language style. Participants who reasoned about either the style of writing, grammatical errors, the tone, or the choice of words were primarily correct about their rating. AI-generated personas appear to use a ``more robotic'' language and write texts that read differently than human-crafted ones. Additionally, human-crafted personas include grammatical errors that are primarily interpreted correctly as a reason to rate a persona as human-crafted. However, some participants still expect AI-generated personas to be less correct sometimes. Furthermore, only a portion of the participants reported grammatical errors as one of the reasons for some of their decisions. At the same time, the choice of words and tone of the language were the main aspects of the writing style.

Participants referred frequently to the personal details and emotional content of human-crafted personas. AI-generated personas were often rated correctly with the reason for including ``unnecessary'' details that do not add information to the persona's personality. In our study, participants expected AI to be less emotional and more general than personal. Thus, AI-generated personas that included personal information were sometimes rated as human-crafted.
AI-generated personas sometimes describe the fictional person's free-time activities in a way perceived as unusual by participants. These fictional persons were described as enjoying doing only things in their free time associated with their work.

Consistency of the given information was another frequent theme in participants' reasoning. Several participants expected human-crafted personas to be more consistent. Since the quantitative results for consistency showed a significantly higher consistency in AI-generated personas, AI can mislead participants due to its quality regarding specific aspects. 

The quantitative results showed higher stereotypically in the AI-generated personas, which was also a frequently mentioned reason for rating a persona as AI-generated. Participants described many AI-generated personas as generic and a stereotype of a person. However, some AI-personas that were described as stereotypical were also perceived as realistic. Some participants seem to recognize stereotypical personas as realistic since the stereotype exists. 

In general, participants argued that personas were human-crafted using a depiction of a person who seems to exist. As shown in the quantitative results, there was no significant difference between AI-generated and human-crafted personas regarding realism. In addition to the argument of personas' likeability and relatability, participants rated personas wrongly as human-crafted if they perceived them to fulfill precisely those aspects.
Likable, relatable, and realistic AI-generated personas were often rated as human-crafted.

The AI-generated personas in our study were often depicted in a more positive light. Participants argued that this purely positive depiction appears unrealistic and rated personas as AI-generated whenever they noticed a description that is only from a positive point of view. The human-crafted personas that talked about the weaknesses of their fictional characters were rated as human-crafted. Thus, a humane persona seems positive and open about strengths and weaknesses.

Our participants pointed out that AI-generated personas often describe the attitude toward technology as less critical and more open. Whenever participants noticed that a persona was critically approaching technology, they rated them human-crafted. This implies the expectation of a bias against technology in AI-generated personas that participants expect. Our participants argued that AI would not describe technology as something negative or something to be careful about. \textbf{We answer RQ2.2:}

\begin{flushleft}   
\setlength{\fboxsep}{10pt}
\setlength{\fboxrule}{2pt}
\fcolorbox{gray!60}{white}{%
    \parbox{.89\columnwidth}{%
    \textbf{RQ2.2:} Participants generally described the human-crafted personas as more realistic, consistent, critical, emotional, and personal. In contrast, AI-generated personas were expected to appear more robotic, include unnecessary details, and rely on overly generic and positive stereotypes. The writing and language style of a persona description, in particular, played a key role in shaping participants' perceptions, with human-crafted personas being seen as more authentic and relatable.
    }
}
\end{flushleft}

\subsection{Generalizability of Findings}

This study used ten personas created by HCI experts who are familiar with the concept but do not regularly design personas. We then evaluated these personas with participants who are not trained in persona design. Although the study’s conditions limit how broadly we can generalize the findings, especially when compared to personas crafted by experts after real user interviews, we believe the results highlight important concerns for more common, less ideal scenarios.

With the growing use of large language models to generate data that typically requires significant human effort, there’s a risk that laypersons might accept AI-generated personas at face value. Because these personas often appear well-written, users may assume they are good enough and skip the deeper design process altogether. Prior studies—and our findings—show that AI-generated personas can seem polished. However, we also show that they lack depth, diversity, and critical realism. If the superficial qualities lead non-experts to rely on them instead of engaging in thoughtful design, the effectiveness of persona-based software development could be seriously undermined.

\subsection{Limitations and Future Work}
This study employed a set of human-crafted personas developed by ten experts in the field of Human-Computer Interaction (HCI). These personas may reflect a realistic example, but are limited in number. Since this study was conducted with online participants, we decided to limit both persona sets to ten so as not to overwhelm the participants. Future work might investigate different aspects of personas more deeply, further to explore the differences between human-crafted and AI-generated personas. Furthermore, we focused on very general personas that were neither designed by routinized design experts nor for a particular use case. While our study aimed at the potential threat to the quality of fast persona generation, it still seems like LLMs could enhance the professional workflows of persona designers. Thus, another study with a main focus on experts can bring further insights.

Participants mentioned the grammatical errors of some human-crafted personas. This could imply that human-crafted personas are easily identified whenever mistakes due to human error are present. However, the recognized grammar mistakes were only present in four personas and interpreted in three cases as human-crafted features. Therefore, the results do not suggest grammatical errors as the main reason why people were able to distinguish the personas. We eliminated all typographical errors before the human-crafted personas were included in the survey to avoid the effect of obvious mistakes on the judgment of our participants. To assess the specific effect of grammatical errors in text on the perception of human users and their awareness of the usage of AI, future research might come up with studies specifically designed for that purpose.

The constructs used to rate the personas in this study are based on related literature. However, we had to limit the number of constructs to those applicable to only text-based personas rated by participants without pre-screening for development experience. The decision to sample participants without a background in computer science, human-centered design, or development was essential to investigating the general perception of human beings as personas. However, experiences in one of the mentioned disciplines might influence participants' perceptions. 

A limitation of this study is the lack of direct evaluation of personas in active design scenarios with professional designers. While our approach provides valuable insights into how personas are perceived regarding realism, consistency, and informativeness, it does not measure their impact on design outcomes or creativity. Future research should involve designers applying these personas in user-centered design tasks to assess their practical utility and effectiveness. Additionally, our findings may be influenced by the participants’ varying levels of familiarity with persona use, which could affect subjective judgments.

Finally, one limitation of our persona generation process was using relatively simple prompts without providing detailed examples or context to guide the language style. While this choice reflects a baseline scenario where users leverage LLMs with minimal customization, it may have contributed to the more robotic tone noted by participants. Future studies should investigate the impact of more elaborate prompt engineering, including few-shot prompting and tailored contextual instructions, to evaluate whether these techniques can produce more human-like and engaging personas.

\section{Conclusion}
This study investigated whether AI-generated personas can be distinguished from human-crafted personas. We collected ten human-crafted personas from ten HCI experts who had no deep experience in persona creation. We asked participants to explain how they perceive these human-crafted personas compared to ten AI-generated personas. We showed that participants could distinguish between the human-crafted and AI-generated personas. In a survey study, we showed that AI-generated personas, compared to personas from novices, are more informative, consistent, and clear, even though their writing and language style were described as somewhat robotic and unnatural. The AI-generated personas were significantly more positive and stereotypical, which was perceived as unrealistic and generic. Overall, we conclude that LLMs can generate personas that meet many quality aspects but also include stereotypical depictions of characters. This can lead to a bias that motivates the usage of personas that might be interesting from a surface-level general quality-based perspective but will retain stereotypes that threaten diverse requirements for engineering and development and, thus, quality.
\begin{acks}
This work is supported by the German Research Foundation (DFG), CRC 1404: ``FONDA: Foundations of Workflows for Large-Scale Scientific Data Analysis'' (Project-ID 414984028). 
\end{acks}

\bibliographystyle{ACM-Reference-Format}
\bibliography{main}

\end{document}